%% file: main_jatis_rev.tex
\documentclass[12pt]{spieman}

\usepackage{rotating}
\usepackage{amsmath}
\usepackage{amssymb}
\usepackage{graphicx}
\usepackage{wrapfig}
\usepackage{ctable}
\usepackage{tocloft}

\usepackage{lineno}


\hypersetup{linkcolor=blue,citecolor=blue,filecolor=cyan,urlcolor=blue}

\newcommand{\unit}[2]{\ensuremath{\textrm{#1}^{#2}}}

\newcommand{\uas}{\ensuremath{\mu\textrm{as}}}
\newcommand{\at}{\ensuremath{@}}

\newcommand{\farcs}{$.\!\!^{\prime\prime}$}
\newcommand{\msun}{M$_\odot$}

\newcommand{\upenn}{Department of Physics \& Astronomy, University of Pennsylvania, 209 South 33rd Street, Philadelphia, PA 19104, USA}
\newcommand{\cca}{Center for Computational Astrophysics, Flatiron Institute, 162 Fifth Avenue, New York, New York 10010. USA}

\newcommand{\stsci}{Space Telescope Science Institute, 3700 San Martin Drive, Baltimore, MD 21218, USA}
\newcommand{\tapir}{TAPIR, California Institute of Technology, 1200 E California Blvd, Pasadena, CA 91125, USA}
\newcommand{\berkeley}{Department of Astronomy, University of California, Berkeley, 501 Campbell Hall, Berkeley, CA 94720, USA}
\newcommand{\princeton}{Department of Astrophysical Sciences, Princeton University, Peyton Hall, Princeton, NJ 08544, USA}
\newcommand{\gsfc}{Astrophysics Science Division, NASA Goddard Space Flight Center, Greenbelt, MD 20771, USA}
\newcommand{\jpl}{Jet Propulsion Laboratory, California Institute of Technology, 4800 Oak Grove Drive, Pasadena, CA 91109, USA}
\newcommand{\osu}{Department of Astronomy, Ohio State University, 140 W. 18th Ave., Columbus, OH 43210, USA}
\newcommand{\ucsc}{University of California Santa Cruz, 1156 High Street, Santa Cruz, CA 95060, USA}

\title{Astrometry with the Wide-Field InfraRed Space Telescope}

\author[$\dagger$]{The WFIRST Astrometry Working Group}
\author[a,b,c,*]{Robyn E. Sanderson} 
\author[d]{Andrea Bellini} 
\author[d]{Stefano Casertano}
\author[e]{Jessica R. Lu}
\author[f]{Peter Melchior}
\author[d]{Mattia Libralato}
\author[g]{David Bennett}
\author[h]{Michael Shao}
\author[h]{Jason Rhodes}
\author[d]{Sangmo Tony Sohn}
\author[g]{Sangeeta Malhotra}
\author[i]{Scott Gaudi}
\author[d]{S. Michael Fall} 
\author[d]{Ed Nelan}
\author[j]{Puragra Guhathakurta}
\author[d]{Jay Anderson}
\author[b]{Shirley Ho}

\affil[a]{\upenn}
\affil[b]{\cca}
\affil[c]{\tapir}
\affil[d]{\stsci}
\affil[e]{\berkeley}
\affil[f]{\princeton}
\affil[g]{\gsfc}
\affil[h]{\jpl}
\affil[i]{\osu}
\affil[j]{\ucsc}

\cftpagenumbersoff{figure}
\cftpagenumbersoff{table}

\begin{document}

\maketitle 

\begin{abstract}
The Wide-Field InfraRed Space Telescope (WFIRST) will be capable of
delivering precise astrometry for faint sources over the enormous
field of view of its main camera, the Wide-Field Imager (WFI). This
unprecedented combination will be transformative for the many
scientific questions that require precise positions, distances, and
velocities of stars. We describe the expectations for the astrometric
precision of the WFIRST WFI in different scenarios, illustrate how a
broad range of science cases will see significant advances with such
data, and identify aspects of WFIRST's design where small adjustments
could greatly improve its power as an astrometric instrument.
\end{abstract}


{\noindent
  \footnotesize{$\mathbf{\dagger}$}\linkable{https://outerspace.stsci.edu/display/FWG/Astrometry}}\\ {\noindent
  \footnotesize\textbf{*}Robyn E. Sanderson,
  \linkable{robynes@sas.upenn.edu} }

\section{Introduction}


The wide field of view and stable, sharp images delivered by the
Wide-Field Imager (WFI) planned for the Wide-Field InfraRed Space
Telescope\footnote{\url{http://www.stsci.edu/wfirst/observatory}} (WFIRST; Ref.~\citenum{2013arXiv1305.5425S}) make it an excellent instrument for astrometry, one
of five major discovery areas identified in the 2010 Decadal
Survey. WFIRST has two main advantages over other spacecraft missions:
it can precisely measure very faint stars (complementary to
\textit{Gaia}); and it has a very wide field of view (complementary to
the \emph{Hubble Space Telescope}, \textit{HST}, and the James Webb
Space Telescope, JWST).  Compared to \textit{HST}, WFIRST's wider
field of view with similar image quality will provide many more
astrometric targets in a single image, but also hundreds more anchors
to the astrometric reference frame in any field, including both
background galaxies and stars with precise positions in the
\textit{Gaia} catalog (Refs.~\citenum{gaia1, gaia2a}. In addition,
WFIRST will operate in the infrared (IR), a wavelength regime where the
most precise relative astrometry has so far been achieved with
adaptive optics images from large ground-based telescopes (e.g. 150
\uas\ from Keck (Ref.~\citenum{2008ApJ...689.1044G}).  WFIRST will
provide at least a factor of three improvement in astrometry over the
current state of the art in this wavelength range, while spanning a
field of view thousands of times larger. WFIRST is thus poised to make
major contributions to multiple science topics in which astrometry
plays an important role. In most cases, these contributions can be
achieved without major alterations to the planned mission or
instrument. In this paper, we summarize a few of the many compelling
science cases where WFIRST astrometry could prove transformational,
and then outline the areas where a small investment of attention now
will ensure that WFIRST's impact on this science is significant.

\begin{table*}
\begin{center}
\begin{tabular}{lll}
\hline
\hline
\textbf{Context} & \textbf{Estimated performance} & \textbf{\S} \\
\hline
Single-exposure precision & 0.01 px; 1.1 mas & \ref{ss:1.1} \\
Typical guest-observer program (100 exposures of one field)  & 0.1 mas & \ref{ss:1.1} \\
Absolute astrometry accuracy & 0.1 mas & \ref{sec:3} \\
Relative proper motions derived from High-Latitude Survey & 25 \uas\ \unit{yr}{-1} & \ref{sec:hls-reqs} \\
Relative astrometry, Exoplanet MicroLensing Survey (per image) & 1 mas & \ref{ss:4.2} \\
Relative astrometry, Exoplanet MicroLensing Survey (full survey) & 3--10 \uas\ & \ref{ss:4.2} \\
Spatial scanning, single scan & 10 \uas & \ref{sec:exoplanets} \\
Spatial scanning, multiple exposures & 1 \uas & \ref{sec:exoplanets} \\
Centering on diffraction spikes & 10 \uas & \ref{sec:exoplanets} \\
\hline
\hline
\end{tabular}
\end{center}
\caption{Approximate expected astrometric performance of the WFIRST
  Wide-Field Imager for different types of observations. All estimates
  are for well-exposed point sources (refer to
  Ref.~\citenum{2013arXiv1305.5425S} and Table \ref{tbl:hls-limits} for
  depths of the core survey programs). See the referred sections for
  details about the assumptions leading to each number. These
  estimates are order-of-magnitude only.}
\label{tbl:astro}
\end{table*}

\subsection{Expected astrometric performance of the WFIRST Wide-Field
  Imager}
\label{ss:1.1}

The astrometric performance of the Wide-Field Channel (WFC) in the WFI
on WFIRST will depend on numerous elements, including hardware
characteristics, the stability of the platform, characterization of
the optics and of the detector (down to the individual pixel), the
ability to design observations with the required properties for
reference stars, and calibration of both the point-spread function
(PSF) and of the geometric distortion (GD) of the focal plane. Here we
summarize the assumptions used in this work; Table \ref{tbl:astro}
collects some performance estimates for different types of
observations.

Ref.~\citenum{2015JKAS...48...93G} discusses astrometric science in
the WFIRST Exoplanet MicroLensing (EML) survey field, but makes very
optimistic sensitivity estimates, and ignores potential sources of
systematic errors. For the purpose of this document, \textbf{we assume
  that the single-exposure precision for well-exposed point sources is
  0.01 pixel, or about 1.1 mas;} this is consistent with current
experience on space-based platforms such as \textit{HST}, as long as a
comparable level of calibration activities are carried out
\cite{Hosek:2015}. Depending on the platform stability and the quality
and frequency of calibrations, this precision can be substantially
improved by repeated, dithered observations as
\begin{equation}
\label{eq:astromAccuracy}
\sigma_{\alpha,\delta} \propto \frac{\Delta \eta}{\sqrt{N}}
\end{equation}
where $\Delta$ is the single-exposure point-source precision in
pixels, $\eta$ is the plate scale, and $N$ is the number of
observations.  We assume that a gain by a factor $\sim 10$ (to $0.1$
mas; i.e. 100 exposures) can be achieved for a typical Guest-Observer
(GO) program.  The EML survey, which will obtain many thousands of
observations of each source, has more stringent requirements
(\S~\ref{ss:4.2}). For proper motion (PM) measurements, the achievable
accuracy depends on the single-image precision and the time baseline
$T$ between the first and last images. In the case of $N$ evenly
spaced images:
\begin{equation}
\label{eq:PMAccuracy}
\sigma_{\mu} \propto  \frac{\Delta \eta}{T \sqrt{N}}.
\end{equation}

Improvements in astrometric measurements can also be obtained by
special techniques, such as spatial scanning and centering on
diffraction spikes, described in \S~\ref{sec:exoplanets}, and by
improvements in the pixel-level calibration, as discussed in
\S~\ref{subsec:subpix}.

Thanks to its large field of view (FoV) and the availability of
accurate reference sources from Gaia, each WFC exposure can achieve an
\textit{absolute} positional accuracy of $0.1$ mas or better (see
\S~\ref{sec:3} for details).   Although WFIRST can directly measure only relative parallaxes and proper motions
   within its field of view, the ability to use reference stars in common with Gaia will allow
   parallaxes and proper motions to be converted to an absolute reference 
   frame to a worst-case accuracy of 10 $\mu$as in parallax and 10 $\mu$as $\unit{yr}{-1}$ in proper motion. ~\\

\section{Science with WFIRST Astrometry}
\label{sec:science}

\begin{table*}[t!]
\begin{center}
\begin{tabular}{lp{3.5in}ll}
\hline \hline
\S & \textbf{Science case}&\multicolumn{2}{c}{\textbf{Astrometric precision}}\\
\hline
\ref{ss:2.1} & Motions of dwarf satellites in the Local Group &$2.2\times 10^{-4}$ pixel\,yr$^{-1}$ & 25 \uas\,yr$^{-1}$\\

\ref{ss:2.2} & Motion of stars in the distant MW stellar halo & $\le 2\times 10^{-4}$ pixel\,yr$^{-1}$ & $\le 25\,\uas$\,yr$^{-1}$  \\

\ref{ss:2.3} & Low-mass end of the subhalo mass function & $1.8\times 10^{-4}$ pixel\,yr$^{-1}$ & 20 \uas\,yr$^{-1}$\\

\ref{sec:exoplanets} & Detection \& characterization of exoplanets & $\le 9\times 10^{-5}$ pixel & $\le 10$ \uas\\

\ref{ss:2.5} & Structure of the MW bulge & $\le 9\times 10^{-5}$ pixel & $\le 10$ \uas\\

\ref{ss:2.6} & Star formation in the MW & $\le 4.5\times 10^{-4}$
                                              pixel\,yr$^{-1}$ & $\le 50$ \uas\,yr$^{-1}$\\

\ref{ss:2.7} & Isolated black holes \& neutron stars &$4.5\times 10^{-4}$ pixel & 50 \uas\\

\ref{ss:2.8} & Internal kinematics in GCs & $\lesssim 1.8\times 10^{-4}$ pixel\,yr$^{-1}$ &$\lesssim 20$ \uas\,yr$^{-1}$\\
\hline\hline
\end{tabular}
\caption{Required astrometric precision (in units of both WFI pixels
  and \uas) for the different science cases discussed in \S\ref{sec:science}.}
\label{tab1}
\end{center}
\end{table*}

The science enabled by astrometry with WFIRST spans size scales from
the Local Group to exoplanetary systems, and provides important
contributions to all three astrophysics goals in the NASA Science
Plan. In this section, we survey the range of science topics to which
this instrument can make important contributions. The astrometric
precision needed for each of the following science cases is listed in
Table~\ref{tab1}.

\subsection{Motions of Local Group galaxies}
\label{ss:2.1}

\begin{figure*}[ht]
\centering
\includegraphics[width=\textwidth]{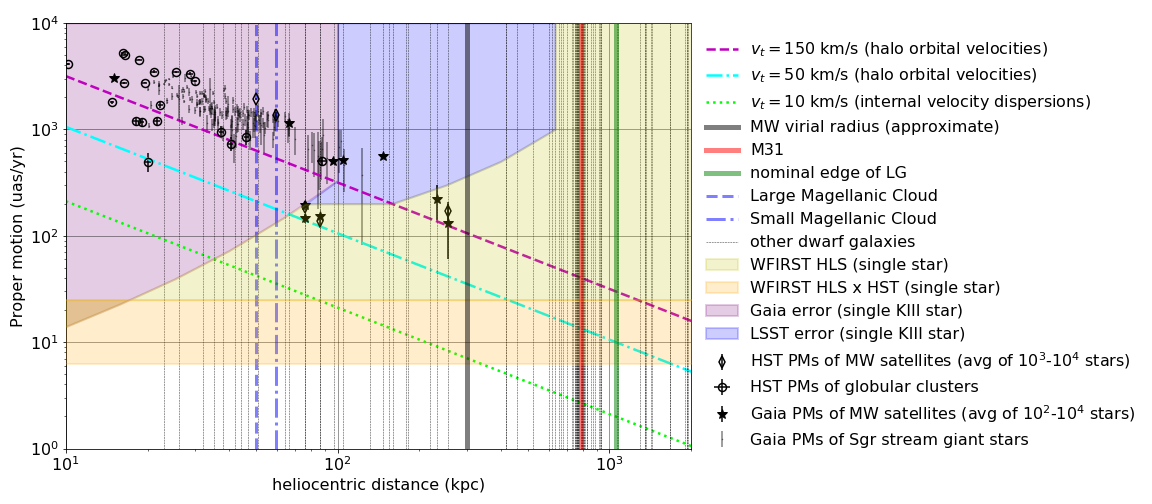} 
\caption{
Proper motions accessible to various current and planned surveys
  and measurements, compared to the PMs corresponding to
  characteristic velocities and distances for Local Group
  objects. Shaded regions show the distances and PMs for single stars accessible to the
  \textit{Gaia} (magenta) and LSST (purple) surveys, compared to the approximate reach of the WFIRST HLS field for bright K giants assuming 15 exposures over 5
  years (yellow) and the additional reach for cross-matches with HST imaging (orange).
  The diagonal lines show the PM associated with several characteristic
  transverse velocities as a function of distance: the typical range of orbital velocities in the Galactic halo (magenta \& cyan) and the typical internal velocity dispersion of a dSph galaxy (green).
  Thick vertical lines mark heliocentric distances to: the Large (blue dashed) and Small (blue dot-dashed) Magellanic Clouds, the edge of the MW halo (grey), M31 (red), and the approximate edge of the LG (green). 
  Grey dotted vertical lines mark heliocentric distances to other dwarf galaxies in the Local Group, including satellites of the MW and M31\cite{2012AJ....144....4M}.
Current PM measurements by HST\cite{Kallivayalil13,2013ApJ...768..139S,2017ApJ...849...93S,Sohn18} and Gaia\cite{2018A&A...616A..12G} for MW globular clusters and satellite galaxies and for individual stars in the Sgr tidal stream\cite{2019ApJ...874..138L} are plotted as black points/symbols.}
\label{fig:PMs}
\end{figure*}

The range and reach of WFIRST astrometry complement and extend
\textit{Gaia} and  Large Synoptic Survey Telescope (LSST) astrometry.  Figure \ref{fig:PMs} compares the
reach of current and planned PM surveys to the PMs corresponding to
known velocities and distances of Local Group (LG) objects. Since the known
orbital and internal velocities refer in almost every case to the
radial component, these are intended only to represent the order of
magnitude one might expect for the PMs (indeed, as in the case of M31,
the orbital PMs may be significantly smaller than that inferred by
radial velocity measurements alone). From this figure it is clear that
to measure PMs of satellites beyond the Milky Way's virial radius will
require better precision than LSST can achieve, at larger distances
(and thus fainter magnitudes). This is the window of opportunity for
WFIRST.

WFIRST astrometry is a crucial component of constraints on the nature
of dark matter (DM) from the orbital and internal PMs of the satellite
galaxies of the Milky Way (MW). The orbits of dwarf satellites can be
used to help map the MW's own DM halo out to its virial
radius and beyond, placing our galaxy into a cosmological context. A
complete knowledge of the MW halo's properties enables tests of dark
matter models through comparisons with predictions from simulations
for its mass and shape, accretion history, and the mass and orbit
distributions of its satellite galaxies. \textit{HST} PMs of dwarf
satellite galaxies, and \textit{Gaia} astrometry in the inner Galaxy,
will both make great strides towards this goal; however, \textit{Gaia}
has insufficient depth, and \textit{HST} insufficient FoV, to reach to
the edge of the MW halo (where the total mass is uncertain to a
  factor of 4; see Ref.~\citenum{2016ApJ...829..108E}) or to obtain internal PM
measurements for many dwarf galaxies (which are crucial to break
velocity degeneracies and understand the small-scale distribution of
DM (see also work by the Gaia Challenge group, \cite{gaia-challenge}
  summarized in Fig 2.2 of Ref.~\citenum{2017arXiv170701348T}.).

\begin{figure}[t!]
\centering
\includegraphics[width=0.5\textwidth]{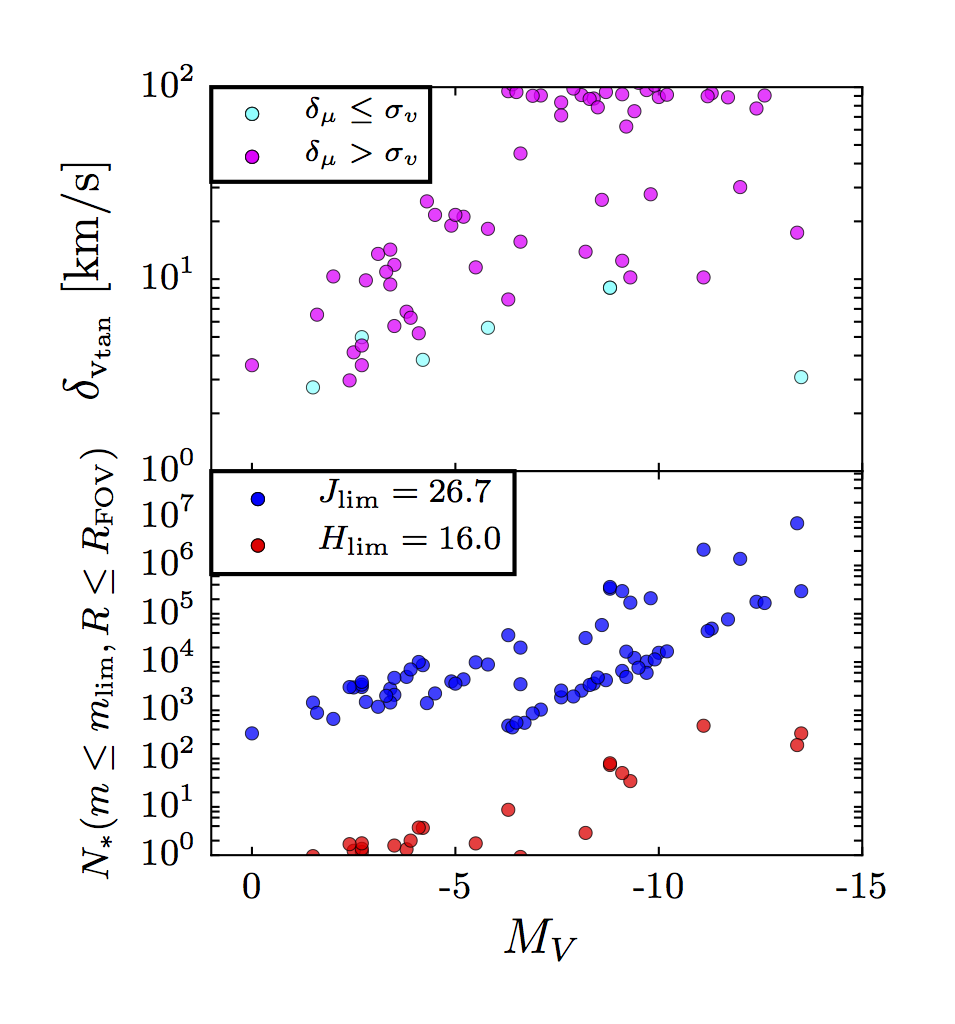}
\caption{{\bf Top:} estimated observational error in tangential
  velocity assuming PM precision of 25 \uas\,yr$^{-1}$. Galaxies in
  cyan have estimated velocity errors ($\delta_{\mu}$) comparable to
  or less than their intrinsic velocity dispersions ($\sigma_v$). {\bf
    Bottom:} number of stars in LG dwarf satellite galaxies
  brighter than the limiting apparent magnitude of the WFIRST HLS ($J<26.7$, blue) and the limit for spatial scanning ($H<16$,
  red). Plot courtesy Matthew Walker.}
\label{fig:wfirst-numbers}
\end{figure}

The HSTPROMO campaign has used the \textit{HST}, which has similar
image quality to that expected for WFIRST, to measure PMs of both
bound objects and stream stars in the MW
(e.g. Refs.~\citenum{2013ApJ...768..139S,2016arXiv161102282S}). In the coming
years \textit{HST} will set a PM baseline for many more distant
satellites \cite{2015arXiv150301785K}:\ by the time WFIRST is ready,
this baseline will be roughly 8--10 years for these satellites, and
far longer for dwarf galaxies with earlier observations. With its
larger FoV and more sensitive detectors, WFIRST should be able to
expand these measurements to more distant galaxies and achieve better
accuracy thanks to the larger number of calibration objects available
(and to the establishment of \textit{Gaia}'s astrometric frame, see
\S~\ref{sec:3}). Figure~\ref{fig:wfirst-numbers} shows the
estimated number of stars in each of the MW's satellites that are
accessible to WFIRST at the depth of the planned High-Latitude Survey
(HLS, $J$$<$26.7, blue) core program, as well as for a possible spatial-scanning mode on
WFIRST ($H$$<$16, red). The upper panel of the figure shows the
expected tangential velocity error for each dwarf assuming PM
precisions of 25 \uas\,yr$^{-1}$. Internal PM dispersions, as was
recently done for the Large Magellanic Cloud with \textit{HST}
\cite{2014ApJ...781..121V}, are reachable with WFIRST for the
galaxies shown in cyan. These include three ultra-faint galaxies
(Segue I, Draco, and Ursa Minor) that are sufficiently DM-dominated to
distinguish between cold DM models (which predict a cuspy
inner mass profile) and warm or ``fuzzy'' DM models (which predict
cored inner profiles). Draco and Ursa Minor each have $\sim$80 stars
at $H$$<$16 and their internal velocity dispersions are marginally
resolved even at 25\,\uas\,yr$^{-1}$, making them particularly good
cases for spatial scanning to improve the internal PM accuracies by an
order of magnitude.

\subsection{Motions of stars in the distant MW halo}
\label{ss:2.2}

Besides dwarf galaxies, the MW's halo contains the tidally disrupted
remains of previously accreted galaxies, known as tidal streams. We
expect that tidal debris should extend to at least the virial radius
of the MW \cite{2017MNRAS.470.5014S}, but currently the most distant
MW halo star known is an M giant at around 250 kpc
\cite{2014ApJ...790L...5B}. The most distant known populations with
statistical samples of stars (BHB, RR~Lyr, and M giant stars) extend
to around 150 kpc (half the virial radius) at the magnitude limit of
current surveys
\cite{2012MNRAS.425.2840D,2014AJ....147...76B,2017ApJ...844L...4S},
but the WFIRST HLS fiducial depth will reach to the MW's virial radius
down to the main-sequence turnoff. The orbits of distant stars probe
the extent and total mass of the MW dark halo;\ they also represent a
unique population of recently accreted small galaxies. With the HLS's
projected depth, the transition between the MW's and M31's
spheres of influence, and perhaps the splashback radius of the MW
(e.g. Ref.~\citenum{2014ApJ...789....1D}), could also be detected. Proper
motions from WFIRST are crucial to these endeavors, since complete
phase-space information for these stars is the best way to confirm
that stars associated in position at large distances are from the same
progenitor and to connect groups on opposite sides of the galaxy
through their orbits, leading to constraints on the mass profile and
flattening of the Galactic dark halo at large distances
(e.g. Ref.~\citenum{2014MNRAS.437..116B}).

High-velocity stars are another interesting target, whether for GO
observations or as serendipitous objects in repeated survey
fields. These stars' orbits, which have an extremely wide radial
range, can potentially also be used to constrain the overall shape and
mass of the MW DM halo \cite{2015ARA&A..53...15B}.

\begin{figure*}[t!]
\includegraphics[width=\textwidth]{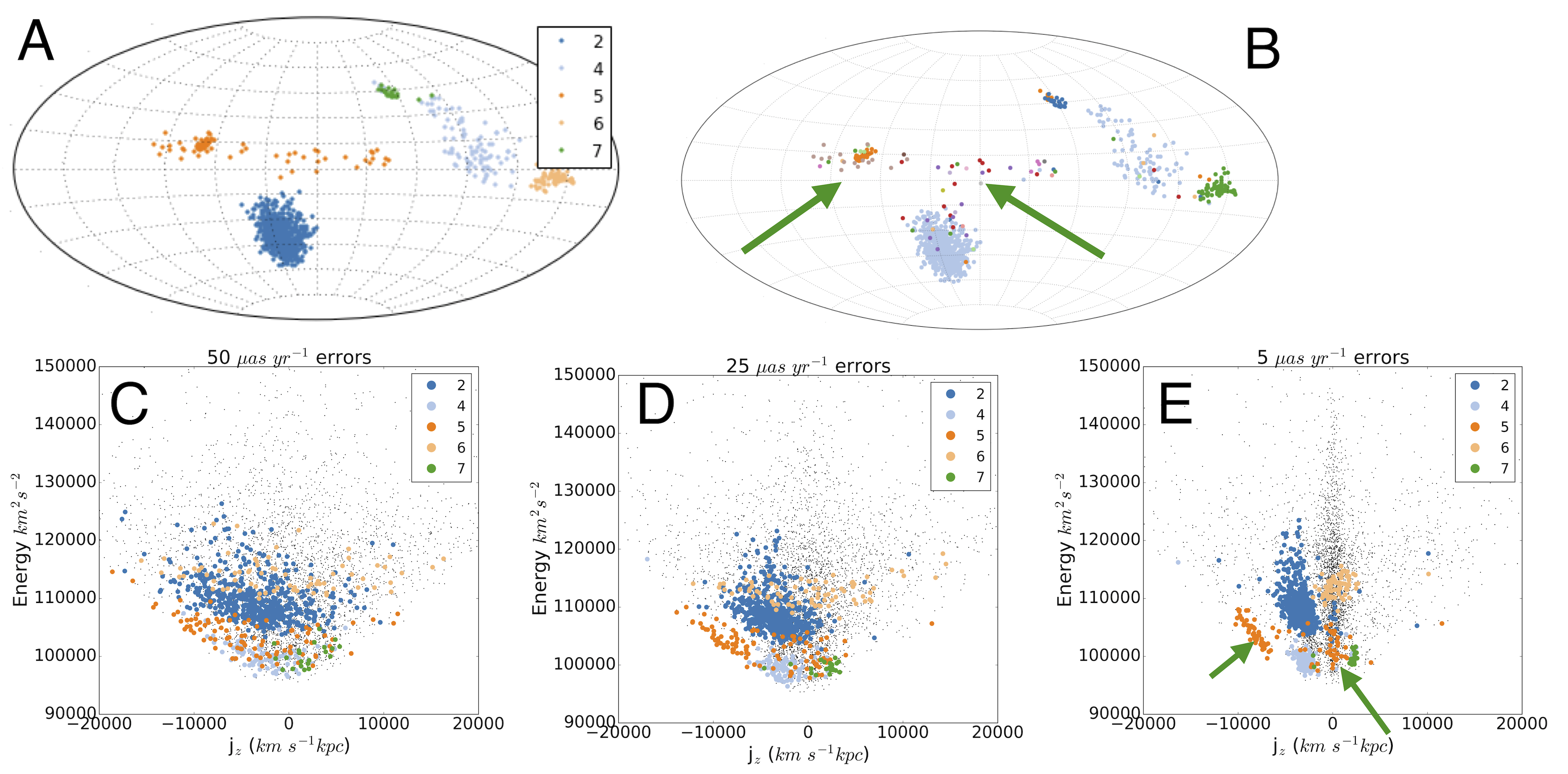}
\caption{Identifying stellar structures in the distant Galactic
  halo. {\bf Panel A:} Groups of stars identified in a mock stellar
  halo \cite{bj05} in the range 100--300 kpc, using sky positions
  (shown) and distances only. {\bf Panel B:} Same stars colored by
  progenitor galaxy. Green arrows highlight the contribution of
  interlopers to group 5 (dark orange) in panel A. {\bf Panels C-E:}
  view of the same groups in energy-angular momentum projection, which
  requires six-dimensional phase space information including PMs. Some
  outliers are already identifiable at 25 \uas\ \unit{yr}{-1} (panel
  D) and structures are clearly distinguishable at 5
  \uas\ \unit{yr}{-1} (panel E; green arrows). Groups 2 and 4 (dark
  and light blue, respectively) in panel A are from the same
  tidally-disrupted progenitor galaxy but are found on opposite sides
  of the sky; with $\le$25 \uas\ \unit{yr}{-1} precision they can be
  associated through orbit integration, reflected in panels D and E by
  their similar values of angular momentum ($j_z$).}
\label{fig:halo_pm_errors}
\end{figure*}

At distances of 100--300 kpc, preliminary work shows that PM precision
of 25 \uas\ \unit{yr}{-1} or better is required to eliminate outliers
in groups and connect structures on opposite sides of the galaxy using
their phase-space positions (Figure \ref{fig:halo_pm_errors}; Secunda
et al. in prep). A similar precision would be needed to identify
high-velocity stars at or near the Galactic escape velocity.

\subsection{Constraining the low-mass end of the subhalo mass
  function}
\label{ss:2.3}


Astrometry from pointed Guest Observer (GO) projects could provide several routes to
understanding the distribution of low-mass substructure in the MW. The
abundance (or lack) of low-mass structure is a key to differentiating
between cold and warm DM models, since cold DM predicts
abundant substructure at small scales while in warm DM the mass
function is cut off at a characteristic scale related to the intrinsic
temperature of the DM particle (and hence to its mass in the case of a
thermal relic).

One route is to search for perturbations to tidal tails from distant
globular clusters (GCs) or dwarf galaxies. Current searches are focused on
quantifying substructure in the spatial distribution of tidal debris
but are limited by knowledge of the MW background/foreground in the
region of the stream as well as by Poisson fluctuations in the star
counts (e.g.~\citenum{2016ApJ...819....1I}) and by the uncertain
dynamical ages of streams, which depend on modeling the orbit
(e.g.~\citenum{2017MNRAS.466..628B}). WFIRST's capability to reach deep
into the stellar main sequence (MS) at these distances will help mitigate
the shot-noise issue. More importantly, obtaining astrometry of fields
including tidal streams would allow superior selection of stream stars
relative to the background/foreground, improving the sensitivity to
density fluctuations and allowing better constraints on the time when
material was first tidally stripped. Streams commonly stretch tens of
degrees over the sky, so WFIRST's large FoV is uniquely well
suited to this application. A thin stream usually has a velocity
dispersion of 1--10 km\,s$^{-1}$, so to provide a useful PM selection
for a stream at 50 kpc would require relative PMs to a precision of
about 20\,\uas\,yr$^{-1}$.

Another possibility is to search for deviations in the apparent
positions of quasars due to strong lensing by dark substructures. For
a distant quasar lensed by a $10^8\ M_{\odot}$ subhalo at 50 kpc, the
Einstein radius of the lens is roughly 20 \uas\ (presuming a singular
isothermal sphere). One route would be to look for the time-dependence
of the lensing around a single quasar:\ for a subhalo moving at 200
km\,s$^{-1}$, the time to cross the Einstein radius is about 10
days. Alternatively, a wide field containing many quasars could be
examined for statistical deviations between exposures taken at
different times (separated by longer than 10 days). Either approach
would require absolute astrometry (i.e. consistent between exposures)
accurate to 20 \uas.

\subsection{Detection and characterization of exoplanets}
\label{sec:exoplanets}

Very accurate PM estimates will make it feasible to search for the
astrometric signature of exoplanets around nearby stars.  For
competitive constraints on exoplanet masses and orbital parameters, an
instantaneous precision of better than 10\,\uas\ is required.  The
best constraints can be achieved for the most nearby stars ($d\lesssim
10\,$pc).  A dedicated GO program that specifically observes those
most promising targets with a flexible schedule is therefore
complementary to the EML.

Because of their close proximity, the target stars are generally very
bright.  This makes high-precision astrometry possible by using one of
two different methods: spatial scanning and diffraction spike
modeling.

\subsubsection{Spatial scanning}

Spatial scanning involves intentionally slewing the spacecraft during
integration to create extended tracks from bright target and reference
stars in the field of interest. This spreads out the signal from each
star over hundreds or thousands of pixels, thereby avoiding saturation
while integrating orders of magnitude more photons, and averaging over
pixel-level artifacts that may significantly affect pointed
observations \cite{2014ApJ...785..161R, Casertano:2016}. Scans in different sky directions can be combined to yield high precision for both coordinates. \textit{HST}
has attained precisions of 20--40\,\uas\ with this technique, limited
in part by the small number of available reference stars and the
variation of the focal plane geometry on the orbital time scale of the
telescope ($\sim 1$ hour).  Because of its larger FoV and more stable
orbit, we expect that the WFIRST WFC will be able to achieve
precisions closer to the noise limit, about 10 \uas\ per exposure. By
combining multiple exposures it will then be possible to achieve a
final relative astrometric accuracy of $\sim 1$ \uas.

Desirable slew rates for spatial scanning are
0.5--$10^{\prime\prime}$\,s$^{-1}$, roughly corresponding to 12--250
pixels per read;\ this is the length of the region over which the
light from each star will be spread within one readout frame. The
fast, non-destructive reads of the WFIRST WFC will allow a clean
separation of the signal accumulated within each pixel from different
sources {\it at different times}, greatly reducing the confusion due
to overlapping trails that has affected applications of this technique
using the Wide-Field Camera 3/Ultraviolet-VISible (WFC3/UVIS) detector on \textit{HST}.  Within the desired range of scanning speeds, it will
be possible to observe unsaturated sources 7 mag brighter than the
pointed-observation saturation limit, or $H_{\rm AB} \sim 4$ mag. The
fastest available scan speed affects primarily the maximum brightness
of the source that can be accomodated;\ slower scan speeds in the
range 0.5--$2^{\prime\prime}$\,s$^{-1}$ can achieve essentially the
same benefits, but with a fainter saturation limit.

Both confusion effects and the signal-to-noise ratio for spatial
scanning observations would benefit more than pointed observations
from obtaining all independent reads for each exposure: unlike pointed
observations, signal does not build up linearly over time in each
pixel, but is deposited there during the narrow time interval in which
a star passes over that pixel.  Extending the interval between
available reads increases both the background accumulated in each
pixel without a corresponding increase in the signal, and the time
interval over which signal from different stars in the same pixel
cannot be cleanly separated.  The availability of intermediate reads
for download is of course subject to mission-level limits on science
telemetry, so the number of reads to download may need to be
determined on a scene- and project-dependent basis.  Finally, spatial
scanning observations will most likely need to be obtained under gyro
control, as the required motion of the spacecraft will quickly exceed
the size of the guiding window.  More details, including error
budgets, will be included in an upcoming white paper (Casertano et
al., in preparation).

\subsubsection{Centering on diffraction spikes}

A second potential strategy for obtaining highly accurate astrometry of very
bright stars involves centering on diffraction spikes.  The approach
is facilitated by the properties of the WFIRST H4RG detectors, which, unlike CCDs, 
do not show ``bleeding'' of excess charges from saturated pixels to
their neighbors (see \S \ref{subsec:subpix}).  Astrometric precisions
of 10 \uas\ or better are achievable with this technique with
integrations of 100\,s for stars with $J=5$ or, making use of the
recently added optical R062 filter, $R=6$
\cite{2017arXiv170800022M}. (Diffraction-spike measurements
  are superior in the short-wavelength range because the diffraction
  spike is sharper. Unlike the core of the WFIRST PSF, the diffraction
  spike is well sampled even in the bluest WFIRST filter given a pixel
  scale of 0.11\,mas.)

As for spatial scanning, measurement accuracy will likely be limited
by systematic uncertainties, in particular the fidelity of corrections
for optical distortion and pixel-level artifacts
(cf.\ \S~\ref{sec:recommendations}).  Thus, performing several
exposures per visit is beneficial and should be able to yield
precisions of 10 \uas\ or better even in the presence of residual
systematics.

\subsubsection{Detection of Earth-mass exoplanets}

These estimates indicate that a dedicated GO program with visits to
target fields separated by months and spread out over the lifetime of
WFIRST could detect Earth-mass exoplanets astrometrically around the most nearby
stars, in some cases even in their respective habitable zones.  In
addition, it can probe Neptune-class planets around more distant stars
and, by adding earlier measurements from \textit{Gaia}, rocky planets
with periods of $>$10 yr.  Such measurements would be strongly
synergistic with radial velocity campaigns, improving the mass
constraints and breaking degeneracies in several orbital parameters
\cite{2000A&AS..145..161P}, and enabling mass estimates of the
direct-imaging exoplanets of the WFIRST coronagraph and possible
starshade occulter programs \cite{2017arXiv170800022M}.

\begin{figure*}[!t]
\begin{center}
\includegraphics[width=\textwidth]{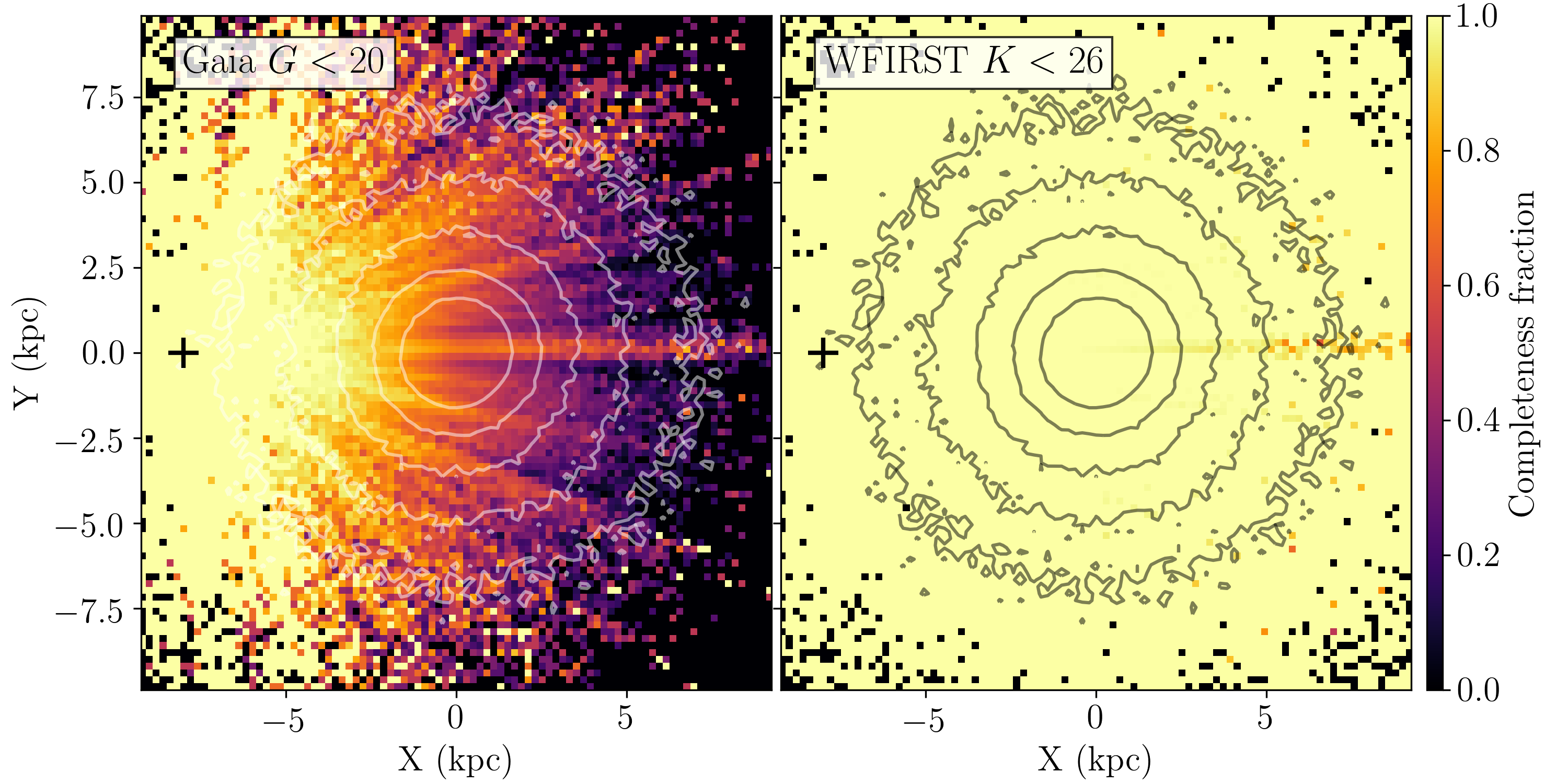}
\end{center}
\caption{\label{fig:gaia_view_of_mw} Simulated completeness of the distribution of red clump stars with distances $|z|<500$ pc from the Galactic plane, based on the Milky-Way-like simulated galaxy in Ref.~\citenum{wetzel16}. The left panel shows stars that \emph{Gaia} would detect in the optical with $G<20$; the right panel those seen by WFIRST in the IR with $K<26$. Grey contours in both panels show the density of the complete distribution on a logarithmic scale; the black cross marks the location of the Sun. The synthetic red-clump catalog was constructed by drawing stars in the range $-0.48<M_{I}<-0.08$, $0.8<V-I<1.4$  \cite{1998ApJ...494L.219P} from the model isochrones in Ref.~\citenum{2008A&A...482..883M}, distributed according to the age and stellar mass density of the simulated star particles \cite{2018arXiv180610564S}. The three-dimensional extinction map in Ref.~\citenum{1998ApJ...500..525S} was interpolated to determine apparent magnitudes and reddening, and \emph{Gaia} $G$ magnitudes were calculated using \texttt{pygaia} \cite{pygaia}.  This simulated view ignores the effects of crowding (which significantly affect \emph{Gaia} in the plane but are anticipated to be relevant for WFIRST only within $\sim0.5$ deg of the Galactic Center; see \S \ref{ss:2.6}) and does not include a prominent Galactic bar (see Ref.~\citenum{RomeroGomez:2015}). \emph{Gaia} will largely be limited to heliocentric distances $<$4 kpc in the plane, while WFIRST can measure parallaxes and velocities of stars well beyond the Galactic Center. }
\end{figure*}

\subsection{Detailed structure of the inner Milky Way}
\label{ss:2.5}

{\it Gaia} will revolutionize our understanding of Milky Way structure
in the outer parts of the Milky Way, including the halo. However, {\em
  Gaia} has a very limited view of the inner Milky Way due to the significant extinction in the Galactic plane at optical wavelengths (Figure
\ref{fig:gaia_view_of_mw} shows an illustrative example using a simulated galaxy) as well as crowding (not accounted for in Fig. \ref{fig:gaia_view_of_mw}). WFIRST will probe significantly deeper
into the inner Milky Way and be less seriously affected by crowding than {\em
  Gaia}, allowing us to map the structure and
kinematics of this region and complement the {\em Gaia} view.  As an example, the EML
survey will obtain precise parallaxes and ultra-precise PMs for over
50 million stars in a small area of the Galactic bulge, enabling a
detailed analysis of their kinematics and density distribution.
Currently, studies of bulge stellar populations are limited by the
quality of the PM and the need to remove foreground disk stars,
typically achieved via kinematic or photometric filters (see, e.g.,
Ref.~\citenum{Clarkson:2008}).  Both are statistical in nature and do not
provide a direct determination of the distance to individual stars.
According to the current requirements, a mission-long astrometric
accuracy of 10\,\uas\ or better (with a stretch goal of 3\,\uas)
should be achieved at $H_{\rm AB} = 21.6 $ ({\bf EML 20};\ see
\S~\ref{ss:4.2}).

\begin{figure}[!t]
\begin{center}
\includegraphics[scale=0.3]{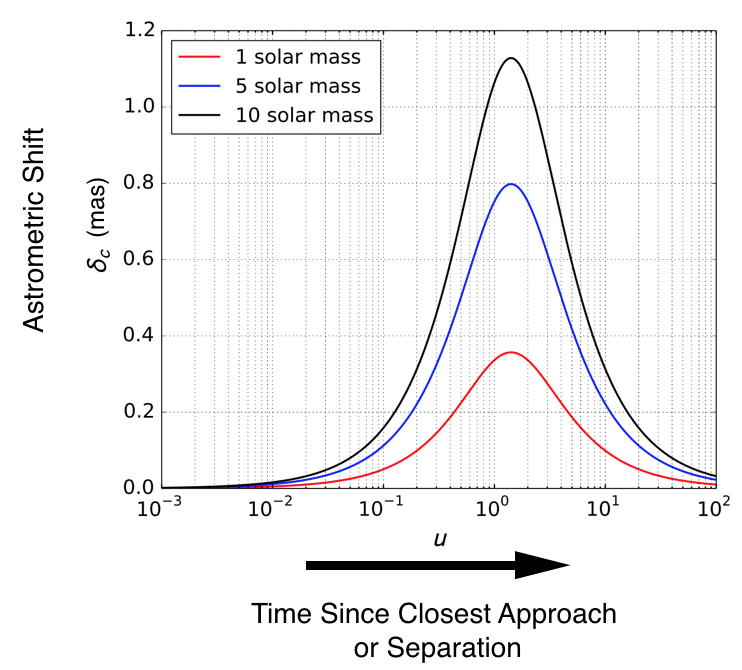}
\end{center}
\caption{\label{fig:microlens} Astrometric shift of a background bulge
  star (source, d$=$8 kpc) lensed by a foreground compact object such
  as a black hole, neutron star, or white dwarf (lens, d$=$4 kpc). The
  astrometric shift changes as a function of the projected source-lens
  separation on the sky, $u$, in units of the Einstein radius. For the
  10\,\msun\ case, the Einstein radius is $\sim$4 mas and the time for
  the source to cross the Einstein ring is typically $>$100 days.  }
\end{figure}

At comparable accuracy in relative parallaxes, distances to individual
stars can be measured to 9\% at the bulge (3\% if the stretch goal is
achieved), and useful distance discrimination should be obtained to
significantly fainter magnitudes.  The two tangential components of
the space velocity can be recovered to the same accuracy (in this
regime, the distance error is dominant over the PM error in deriving
the space velocity).  This information will enable a much cleaner
determination of the kinematics of the bulk of bulge stars in the EML
survey field, and readily identify subgroups of stars---disk or
bulge---with anomalous kinematics.  If depth effects can be accounted
for, the end-of-mission PM accuracy translates to a velocity precision
of $\sim 1$ km\,s$^{-1}$;\ together with the very large number of
stars measured, this will permit a clear component separation of the
spatially overlapping bulge and halo populations (see, e.g.,
Ref.~\citenum{Minniti:2008}), and potentially identifying complex
structures such as the anomalous motions found in the X-shaped regions
of the bulge (see, e.g., Ref.~\citenum{Vasquez:2013}).  In principle,
\textit{Gaia} will achieve comparable precision over all of the bulge,
but only for the bright red giants at $G \sim 15$ or
brighter;\ the uncertainties will be considerably larger ($\sim 2$
orders of magnitude) at \textit{Gaia}'s faint limit.

Regions within 0.5 deg of the Galactic Center will likely suffer from crowding at $H<21$, as has been seen with \textit{HST}-WFC3IR studies of this region\cite{2019ApJ...870...44H}, which would limit the astrometric precision to $>0.5$ mas $\unit{yr}{-1}$ for stars fainter than this. Beyond this region, the stellar density is not typically high enough to impact astrometric precision down to $H<24$. However, the exact determination of how the WFIRST astrometric precision will scale with stellar density, SNR, PSF knowledge, and survey depth will require image-level simulations in the future. 

\subsection{Star Formation in the Milky Way}
\label{ss:2.6}

With the advent of large IR surveys of the Galactic Plane, many
new young star clusters have been identified. The most massive of
these young clusters are ideal laboratories for studies of star and
cluster formation, stellar evolution, and cluster dynamics, but
detailed studies of these regions are hampered by high and spatially
variable extinction, high stellar densities, and confusion with
foreground and background stars. Many of these limitations can be
overcome with the addition of PMs observed in the IR, to
separate out the co-moving cluster members from the contaminating
field population \cite{Stolte:2008,Hosek:2015}. Furthermore,
measurements of the internal velocity structure of star clusters
provide constraints on the unseen stellar population from dynamical
mass measurements, thereby informing cluster evolution models
\cite{Clarkson:2012}.

WFIRST is ideally suited for studies of massive young clusters and
star forming regions in the Milky Way, given its wide FoV at
IR wavelengths, high spatial resolution, and potential for
precise photometry and astrometry. For rapidly moving populations in
the center of the Galaxy, a PM precision of $\sim$0.5 mas\,yr$^{-1}$
per star is needed to separate cluster members from field stars. To
obtain internal velocities or separate clusters in the disk, a PM
precision of 0.05 mas\,yr$^{-1}$ or better is desired. Even higher
astrometric precisions would enable searches for binaries and
higher-order multiples.

An important factor to consider for this science case is that cluster
members span a large range in brightness. The brightest and most
massive cluster members in clusters beyond 4 kpc often have $J = 9$ or
brighter. Careful calibration of persistence, shorter integration
times or possibly narrow-band filters will be needed to reach both
bright, massive members and faint, low-mass members of clusters.

\begin{figure*}[!t]
\centering
\includegraphics[trim=0 0 250 0,clip=true,width=\textwidth]{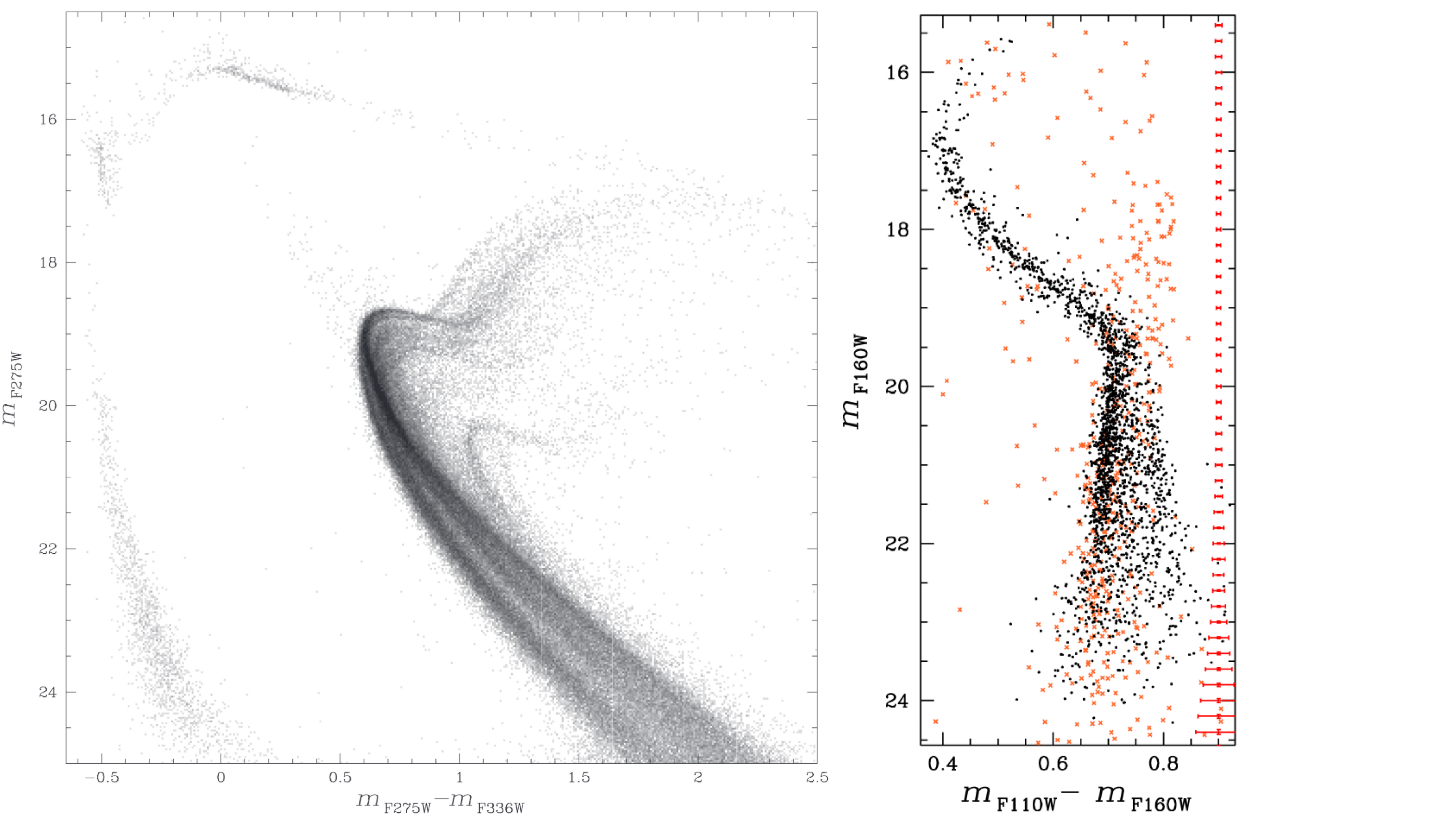}\\
\caption{ \textbf{Left:} The $m_{\rm F275W}$ vs.\ $m_{\rm
    F275W}-m_{\rm F336W}$ CMD of $\omega$~Cen, showing several
  subpopulations of stars in all evolutionary sequences (from
  Ref.~\citenum{2017ApJ...842....6B}). MPs reveal themselves in
  high-precision photometry from the UV to the near IR.
  \textbf{Right:} $m_{\rm F160W}$ vs.\ $m_{\rm F110W} - m_{\rm F160W}$
  CMD of an outer field of $\omega$~Cen, corrected for differential
  reddening. Field stars (orange dots) are identified using proper
  motions.  From Ref.~\citenum{2017MNRAS.469..800M}.}
\label{f:gc1}
\end{figure*}

\subsection{Isolated Black Holes and Neutron Stars}
\label{ss:2.7}

Our Galaxy likely contains $10^7$--$10^8$ stellar mass black holes and
orders of magnitude more neutron stars
\cite{AgolKamionkowski:2002}. Measuring the number and mass
statistics of these stellar remnants will provide important
constraints on the initial stellar mass function, the fate of massive
stars and the initial-final mass relation, the star formation history
of our Galaxy, and the fundamental physics of compact objects. WFIRST
has the ability to find such objects in large numbers through
gravitational microlensing when a background star passes behind the
compact object and is magnified photometrically. However, only the
addition of WFIRST astrometry will enable us to measure the precise
masses of these objects through astrometric microlensing. The apparent
astrometric shift of the background star due to microlensing, which is
proportional to $M^{1/2}$, is $\sim$1 milli-arcsecond for a 10
\msun\ black hole at 4 kpc lensing a background star at 8 kpc (Figure
\ref{fig:microlens}). Thus, the necessary astrometric precision to
detect isolated black holes is $<$150 \uas; a factor of 2-3 better
precision would also allow the detection of neutron stars.

While \textit{Gaia} or ground-based adaptive optics systems may detect
one or a few isolated black holes \cite{Lu:2016}, WFIRST's IR
capabilities and monitoring of the Galactic Center and bulge fields
will yield the much larger samples needed to precisely measure the
black hole and neutron star mass function and
multiplicity. Microlensing by massive objects typically has long
timescales, with Einstein crossing times $>$100 days for black holes,
so WFIRST astrometry should be stable on these timescales; i.e.,
routinely calibrated on sky if possible.

\subsection{Globular clusters}
\label{ss:2.8}

In the last decade, a wealth of revolutionary studies have
dramatically changed the traditional view of globular clusters (GCs)
as the best examples of ``simple stellar populations:'' stars with the
same age and chemical composition. The presence of multiple stellar
populations (MPs) in GCs has been widely established along all the
stellar evolutionary phases (e.g., Ref.~\citenum{2009A&A...505..117C,
  2015AJ....149...91P} and references therein): spectroscopic studies
have found significant star-to-star variation in light elements (e.g.,
Ref.~\citenum{gratton12} and references therein), while high-precision
photometry, mostly from \textit{HST} data, has clearly revealed the
presence of distinct sequences in color-magnitude diagrams (CMDs) at
all wavelengths (e.g., Refs.~\citenum{2012ApJ...754L..34M} and \citenum{2015AJ....149...91P}; see also
Fig.~\ref{f:gc1} of Ref.~\citenum{2017ApJ...844..164B}). Several GCs have also shown the presence of
significantly He-enhanced subpopulations (e.g.,
Ref.~\citenum{2005ApJ...621..777P, 2007ApJ...661L..53P}) and even
subpopulations with distinct iron content in a few cases like
$\omega$~Cen, M22, Terzan~5, M54, NGC~5824, and M2 \cite{1995ApJ...447..680N,2010ApJ...722.1373J,2010A&A...520A..95C,2015MNRAS.447..927M}. These
observational findings present formidable challenges for theories of
the formation and evolution of GCs, and have inaugurated a new era in
GC research in which understanding how multiple stellar systems form
and evolve is not just the curious study of an anomaly, but a
fundamental key to understanding star formation.

Measuring the PMs of stars in GCs is the most effective way to
constrain the structure, formation, and dynamical evolution of these
ancient stellar systems and, in turn, that of the Milky Way itself.
High-precision \textit{HST} astrometry of GCs is now becoming
available for a large number of objects (e.g.,
Ref.~\citenum{2014ApJ...797..115B}), but current PM catalogs are limited by
the small FoV of \textit{HST}, either to the innermost few
arcminutes or to pencil-beam locations in the outskirts.  While most
dynamical interactions do happen in the center of GCs, answering many
outstanding questions will require high-precision PMs of faint cluster
stars over wide fields, for which WFIRST is by far the best tool. Here
we discuss a few examples of such investigations.

\begin{figure*}[t!]
\centering
\includegraphics[width=\textwidth]{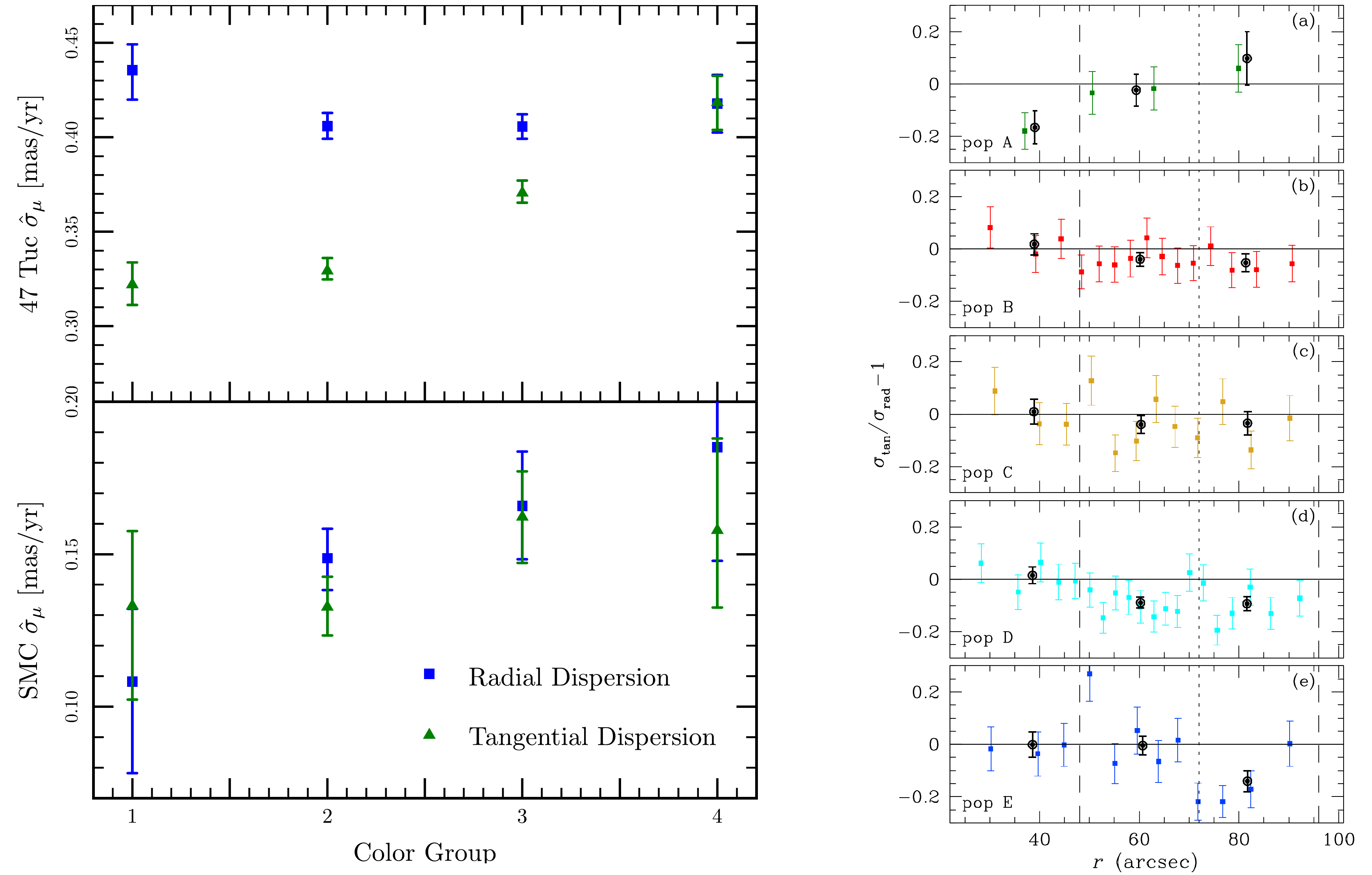}
\caption{ \textbf{Left:} Radial (blue) and tangential (green) PM
  dispersions as a function of color for the bluest (left) and reddest
  (right) parts of the MS of 47~Tuc (top) and of the Small Magellanic Cloud
  (Bottom). (From Ref.~\citenum{2013ApJ...771L..15R}). \textbf{Right:}
  Deviation from tangential-to-radial isotropy (horizontal line) for
  the 5 subpopulations in NGC~2808. Vertical lines mark the locations
  of $r_{\rm h}$, 1.5$\times r_{\rm h}$, and 2$\times r_{\rm h}$ (From
  Ref.~\citenum{2015ApJ...810L..13B}).}
\label{f:gc2}
\end{figure*}

\subsubsection{Multiple-population internal kinematics}

The PM-based kinematic properties of MPs have so far been
characterized for only three GCs: 47~Tuc \cite{2013ApJ...771L..15R}, $\omega$~Cen
\cite{2010ApJ...710.1032A}, and NGC~2808
\cite{2015ApJ...810L..13B}. The short two-body relaxation
timescale in the inner regions of these clusters, where most
observations have so far been focused, implies that any initial
differences in the kinematic properties of different stellar
populations have likely been erased. The cluster outskirts, however,
have much longer relaxation timescales and could still retain fossil
kinematic information about the early stages of cluster evolution
(e.g., Ref.~\citenum{2008MNRAS.391..825D}). The outer regions can thus
provide a wealth of information and constraints on the formation and
early dynamics of MP clusters, on the subsequent long-term dynamical
evolution driven by two-body relaxation, and on the role of the
Galactic tidal field in the outskirts of clusters. For instance, it
has been shown that second-generation stars in 47 Tuc
\cite{2013ApJ...771L..15R}, NGC~2808 \cite{2015ApJ...810L..13B},
%
and $\omega$~Cen \cite{2018ApJ...853...86B}
are characterized by an increasing
radial anisotropy in the outer regions with respect to
first-generation stars (see Fig.~\ref{f:gc2}).  Even further out, at
distances approaching the tidal radius, the effects of the external
tidal field are expected to lead to a more isotropic velocity
distribution.

Both WFIRST's wide FoV and its improved sensitivity will
make revolutionary steps forward in understanding the initial
differences in MPs if sufficient PM accuracy can be achieved. Due to
mass segregation, the most abundant stars in the outskirts of GCs are
low-mass, faint main-sequence objects.  \textit{Gaia} can only measure
stars as faint as the turn-off region in most clusters, and therefore
will not be able to provide enough statistics to properly characterize
the kinematics of the outer cluster regions. The expected internal
velocity dispersion of cluster stars near the tidal radius is of the
order of $\lesssim$ 1--3\,km\,s$^{-1}$, even for the most massive
clusters. The PM error adds in quadrature, so it should be less than
half the intrinsic velocity dispersion, i.e., $\lesssim
1$\,km\,s$^{-1}$, in order to measure dispersions in cluster
outskirts. At the typical distance of Galactic GCs, $\sim$10 kpc, this
translates into PM errors of the order of
$\lesssim$20\,\uas\,yr$^{-1}$.

\subsubsection{Energy equipartition}

\begin{figure}[!t]
\centering
\includegraphics[width=\columnwidth]{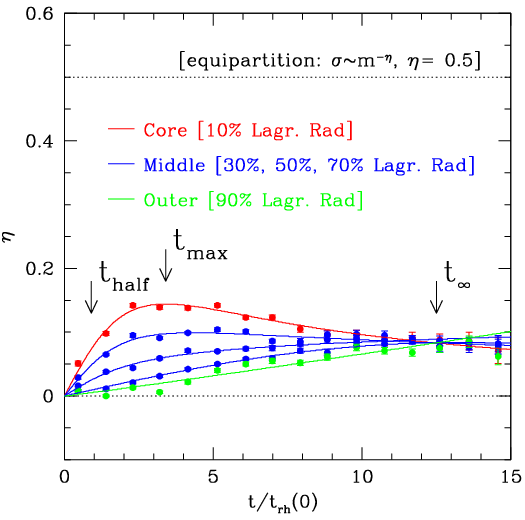}
\caption{Time evolution of the energy equipartition indicator $\eta$
  for single main-sequence stars in N-body simulations, from
  Ref.~\citenum{2013MNRAS.435.3272T}. The time along the abscissa is
  expressed in units of the initial half-mass relaxation time $t_{\rm
    rh}$(0). Complete energy equipartition ($\eta$=0.5; dotted line)
  is never attained, confirming previous investigations based on
  stability analysis.}
\label{f:gc4}
\end{figure}

It is widely assumed that GCs evolve towards a state of energy
equipartition over many two-body relaxation times, so that the
velocity dispersion of an evolved cluster should scale with stellar
mass as $\sigma\propto m^{-\eta}$, with $\eta=0.5$. Recently,
\cite{2013MNRAS.435.3272T} used direct N-body simulations with a
variety of realistic initial mass functions and initial conditions to
show that this scenario is not correct (see also
Ref.~\citenum{2016MNRAS.458.3644B}). None of these simulated systems
reaches a state close to equipartition: instead, over sufficiently
long timescales the mass-velocity dispersion relation converges to the
value $\eta_{\infty}\sim0.08$ as a consequence of the Spritzer
instability (see Fig.~\ref{f:gc4}). 

%
%
These intriguing results have just started to be observationally
tested (e.g., Refs.\ \citenum{2018ApJ...853...86B}, \citenum{2018ApJ...861...99L}).
To measure $\eta$, a wide range of stellar masses must be
probed. Again, this task is out of reach for \textit{Gaia} because of
its relatively bright magnitude limit, but WFIRST will easily measure
high-precision PMs down to the hydrogen-burning limit (HBL;
$\sim$0.08$ M_{\odot}$) and out to the tidal radius, thus constraining
both the current state of energy equipartition in a cluster and its
past dynamical evolution. As in the previous case, PM errors of the
order of $\lesssim$20\,\uas\,yr$^{-1}$ are needed.

\subsubsection{The hydrogen-burning limit and the brown-dwarf regime}

\begin{figure}[!t]
\centering
\includegraphics[width=\columnwidth]{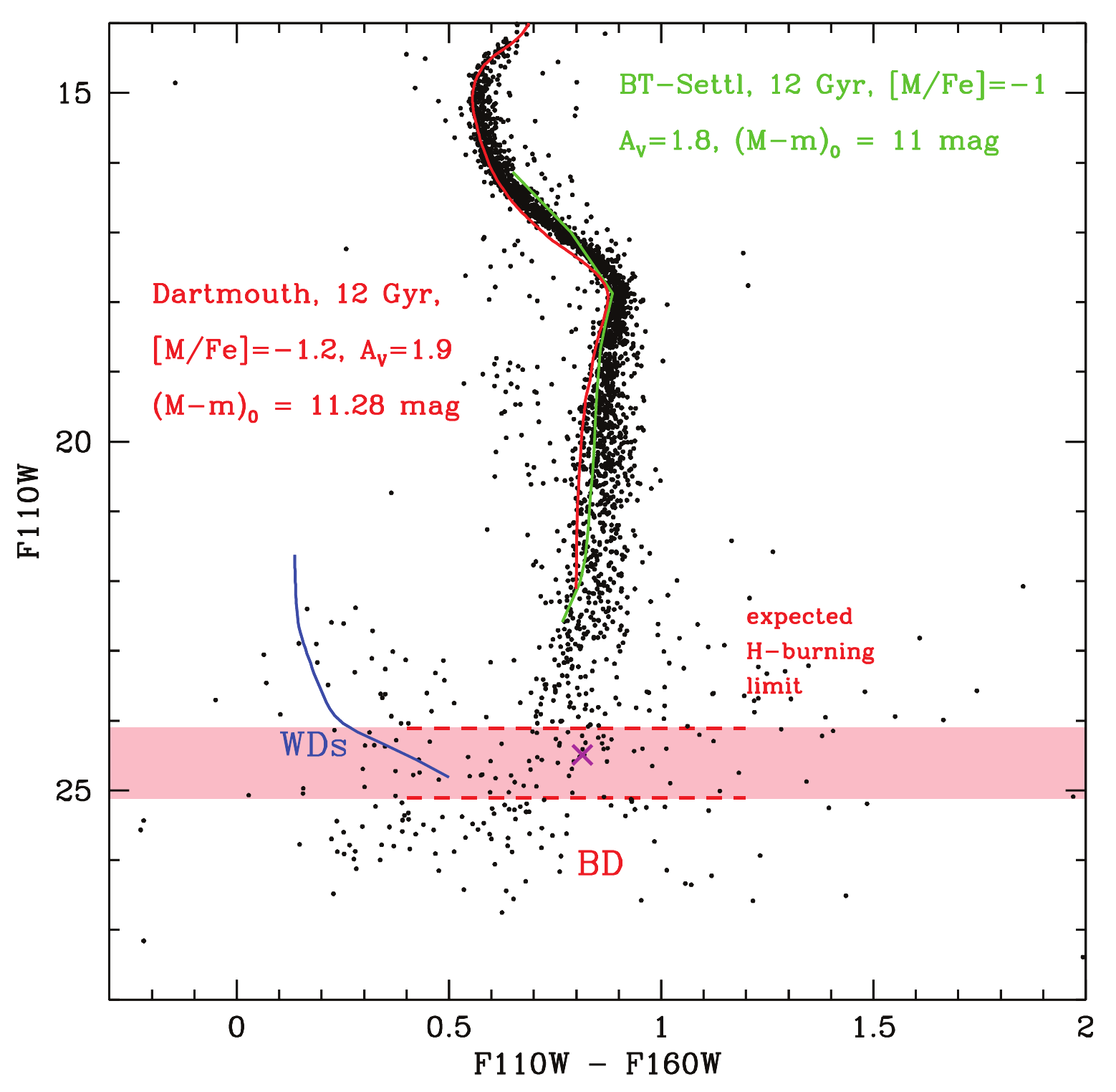}
\caption{Deep near IR CMD of the GC M4, from Ref.~\citenum{2016ApJ...817...48D}.
  The white-dwarf and brown-dwarf regions are labeled, and low-mass
  stellar models are over-plotted in green and red. The expected end
  of the H-burning sequence is marked with red dashed lines and a
  shaded area.}
\label{f:gc3}
\end{figure}

WFIRST will also make it possible to study the luminosity functions of
GCs beyond the HBL and into the brown-dwarf regime.  Close to the HBL,
old stars show a huge difference in luminosity for a small difference
in mass, resulting in a plunge of the luminosity function toward zero
for stars with masses just above this limit.  Stars in GCs are
homogeneous in age, distance, and chemical composition (within the
same subpopulation), so at the typical GC age of $>10$ Gyr, stars with
masses below the HBL will have faded by several magnitudes relative to
those above it, creating a virtual cutoff in the luminosity function
(e.g., Refs.~\citenum{2001ApJ...560L..75B, 2016ApJ...817...48D}; see also
Fig.~\ref{f:gc3}).

The best place to observe the properties of stars approaching the HBL
is once again outside a cluster's core region, where contamination by
light from much brighter red-giant-branch stars is negligible.  The
brown-dwarf regime in GCs is unexplored ground, so many new and
intriguing discoveries may be waiting for WFIRST.  Due to the
relatively low number density of low-mass MS and brown-dwarf stars in
the cluster outskirts, WFIRST is the perfect astronomical tool for
these investigations as well. Proper-motion-based cluster-field
separation is needed to create clean samples of cluster members; PM
errors of the order of a few tenths of mas\,yr$^{-1}$ might be
sufficient to separate cluster stars from field stars, while errors
one order of magnitude smaller would also enable studies of the
internal kinematics of cluster stars in these lowest-mass regimes.

\section{WFIRST absolute astrometric performance}
\label{sec:3}

High-precision absolute astrometric measurements with \textit{HST}
have typically been based on the positions of well-measured background
galaxies within the FoV, but a new major improvement
in absolute astrometry measurements is imminent. When WFIRST begins
observing the universe, the European Space Agency (ESA) mission \textit{Gaia} will already be
complete, providing absolute astrometric positions of unprecedented
accuracy everywhere on the sky. WFIRST will be able to make use of Gaia's absolute astrometric reference frame to convert its relative astrometry to absolute astrometry.

In broad terms, there are two methods to determine the absolute
astrometric accuracy of a scientific observation. Initial
astrometric positions can be obtained from the telescope pointing
information using guide star data (``a priori''). This information includes the
celestial coordinates of the guide stars (GSs) and the locations of
the scientific instruments relative to the GSs in the focal plane of
the telescope.  These positions can be refined based on information
available after an observation is made, namely the positions of all
sources with accurate coordinates in external catalogs (``a posteriori''). Although there will be a minimum of four guide stars, perhaps as many as 18, used to guide WFIRST observations, many more fainter stars within
each exposure can be used to improve the absolute astrometric
precision with the a posteriori method \cite{bellini17b}.

 According to WFIRST's design and operations concept, guide stars can be placed on any of the 18 WFI detectors, but at most one GS will be assigned to a given detector. Assuming the nominal $10 < H < 15.6$ bright and faint GS magnitude limits, analysis has shown (Ref. 76) that the Two-Micron All Sky Survey (2MASS) can provide at least one GS candidate for each of the 18 WFI detectors with a probability close to 100\% (and at least 10 detectors will have a GS candidate brighter than $H = 12.6$). While perhaps as few as 4 bright GS ($H < 13$) will suffice for the attitude control system, the fine guidance system (FGS) design supports the use of up to 18 guide stars since each of the detectors will read out a ``guide window'' to keep the readout pattern for all 18 detectors synchronized, even if the guide window does not contain a guide star.   
 For the grism mode,
the whole sky will be available if the faint limit for GSs is
$H_{\rm AB}<14$, but if the faint limit is pushed to $H_{\rm AB}<13$
or even $H_{\rm AB}<12$, then WFIRST will not be able to perform
spectroscopy in a few regions (0.001\% and 3.5\% of the sky,
respectively) around the Galactic poles.

The WFIRST mission will begin operation in the second half of the
2020s, several years after \textit{Gaia} has completed its nominal
5-year mission in \textbf{July} 2019. However, the \textit{Gaia}
mission has already been extended to the end of 2022\footnote{See the ESA press release \url{http://sci.esa.int/director-desk/60943-extended-life-for-esas-science-missions/}} and could in principle be extended up to a total of 5 years beyond its nominal mission based on the depletion rate of consumables and the degradation rate of the main CCD camera.

Near the faint \textit{Gaia} limit ($19<G<20$), PMs in the
\textit{Gaia} catalog will have an end-of-mission error (assuming the
nominal 5-year baseline) of about 0.2--0.3 mas\,yr$^{-1}$ (see
Ref.~\citenum{2012Ap&SS.341...31D}). This translates into a position
uncertainty of about 1.6--2.4 mas (or $\sim 0.015$--0.02
WFIRST WFI pixels) at the start of WFIRST operations, and about twice
as much by year 5. Position errors of this size will have a
significant impact if the goal is to achieve high-precision (to better
than 0.01 pixels) absolute astrometric measurements. In the following,
we provide expected estimates based only on catalog errors. All other
sources of errors (geometric-distortion residuals, centroiding errors,
etc.) are ignored (more in Ref.~\citenum{bellini17b}). Gaia's
extension for another 5 years (the estimated maximum possible) will improve on the following analysis significantly, not only with improved PMs (lowering
uncertainties by a factor $2\sqrt{2}$ compared to the 5-year
baseline), but also by reducing the timespan over which positions must
be extrapolated (another factor of 2 improvement, assuming WFIRST
begins operation in 2025).

Single-epoch GO and Guest-Investigator (GI) observations of a random
location on the sky may have to rely solely on the information
contained in prior astrometric catalogs (in particular \textit{Gaia}'s
catalog) to determine the absolute position of their sources.
Assuming an average per-star positional error of 2 mas (corresponding
\textit{Gaia}'s expected end-of-mission astrometric error for $G_{\rm
  Gaia}=19$, extrapolated to the late 2020s), and ignoring all other
sources of errors (e.g., geometric-distortion or source centroiding
errors), it will be possible to obtain absolute a posteriori positions
to better than $\sim 0.05$ mas (or about $5\times 10^{-4}$ WFI pixels)
over half the sky. In regions with the lowest stellar densities, the
expected absolute position error increases to $\sim 0.1$ mas ($\sim
10^{-3}$ WFI pixels). 

For the planned WFIRST mission surveys (the HLS and EML surveys),
repeated WFIRST observations spanning several years can be used to
improve \textit{Gaia}'s PMs, especially at the faint end, and to
derive absolute positions and PMs for many fainter sources. The
astrometric precision for the planned surveys is expected to
be significantly better than what can be done with \textit{Gaia}
alone, but is difficult to quantify at this time.

For all stars suitable for the a priori method, \textit{Gaia}'s
expected end-of-mission astrometric error ranges between 10 and 80
\uas\,yr$^{-1}$ \cite{2012Ap&SS.341...31D}, but WFIRST will
likely choose the four GSs among the brightest available sources. We
estimate that at least 7--8 GSs in the range $10.0 < H_{\rm 2MASS} <
10.2$ will be available over half the sky. If these stars land on at
least four different WFI chips (a near 100\% chance;
Ref.~\citenum{nelan15}), assuming their average magnitude is $H_{\rm
  2MASS}\sim 10.1$ (corresponding to $G_{\rm Gaia}\sim 13.7$), and
assuming they have a \textit{Gaia}-extrapolated position error of
$\sim$0.15 mas in late 2020s, then the a priori method is expected to
offer absolute position measurements at the 0.075 mas level or better
($7\times 10^{-4}$ WFI pixels) for half the sky.  For the entire sky,
on the other hand, we always expect at least 7--8 GSs within any given
WFI FoV if the faint limit is relaxed to $H_{\rm 2MASS} = 12.4$ (or
roughly $G_{\rm Gaia}=15.8$). This translates into an upper limit for
the expected a priori astrometric error of 0.2 mas (or about $2\times
10^{-3}$ pixels).

The proposed 5-year extension to the \textit{Gaia} mission would
improve WFIRST astrometry substantially.  In this case each
\textit{Gaia} source will have twice as many measurements over twice
the time baseline, providing an increase in precision by a factor of
$2\sqrt{2}$. Moreover, the time between missions, and hence the interval over which WFIRST would need to extrapolate \textit{Gaia}'s positions, would be reduced by 2.5 years, resulting in an
additional factor of $\sim$\textbf{1.4} improvement at WFIRST's first light. 

Finally, the ability to guide the telescope using more than four GSs, optimally one in each detector, provides the means to monitor the stability of WFIRST focal plane solution for all such observations by comparing the observed relative positions of the GSs to their catalogued positions. The ground system can impose the requirement that these GSs have Gaia positions (and parallaxes and proper motions). The GS positions are reported $\sim 6$ times per second, much more frequently than the WFI full frame images, even more so considering that not all frames will be saved to the recorder. Moreover, the Gaia field stars in the full frame images will likely have saturated PSFs after the first few reads. Therefore the GSs provide a unique opportunity to monitor the focal plane solution, which is critical for high accuracy astrometry. If significantly fewer than 18 detectors routinely host a GSs, then the stability of the focal plane solution may need to be accessed using dedicated calibration observations, with the results interpolated to the intervening science visits. 

The uncertainty of the conversion of \textit{relative} to \textit{absolute} parallaxes and
   proper motions depends on the number of reference sources and 
   their individual Gaia measurement errors.  In regions of low stellar density - e.g.,
   near the South Galactic Pole but away from NGC 288 - the Gaia DR2 catalog contains about 150 stars
   per WFIRST field of view with Gaia magnitudes $17<G<19$.  These stars have median 
   Gaia DR2 uncertainties of $\sim170\ \mu$as in parallax, and $\sim300\ \mu$as $\unit{yr}{-1}$ per component
   in proper motion.   At the end of the Gaia mission, these uncertainties are expect to drop 
   below 100 $\mu$as and 100 $\mu$as $\unit{yr}{-1}$, respectively,
   allowing a conversion to absolute parallax and proper motion with a worst-case error 
   better than 10 $\mu$as in parallax and 10 $\mu$as $\unit{yr}{-1}$ in proper motion.  Typical performance in areas with higher 
   stellar density will likely be much better.  While systematic issues still exist with 
   Gaia parallaxes and proper motions at the level of a few tens of $\mu$as\cite{2018A&A...616A..17A,2018A&A...616A...2L}, improved calibration 
   and processing will likely reduce these substantially in future releases.

\section{Recommendations}
\label{sec:recommendations}

Here we consider what is most likely to have an effect on the
astrometric performance of the WFI. We highlight areas
where astrometry-specific considerations are especially important and
can add significant extra science capability with little to no extra
cost. Our recommendations are summarized in Table \ref{tbl:recs}.

\begin{table*}[tp]
{\setlength{\extrarowheight}{0.3cm}
\begin{tabular}{p{0.3in}p{1.5in}p{4.2in}}
\hline\hline
 \S & Topic & Recommendation  \\
\hline
\ref{subsec:geodist} & Geometric distortion & This significant
systematic error for astrometry, not currently covered by core science
requirements, should be considered in calibration plans for the
WFI. \\

\ref{subsec:subpix} & Pixel-level effects & Ground-based calibration
should be considered, based on results of current tests by several
labs. A spatial scanning mode would mitigate these effects for bright
stars.  \\

\ref{subsec:color} & Filters \& color dependence & Likely straightforward
to calibrate, but should be aware of systematic effects.\\

\ref{subsec:hyster} &Readout Hysteresis & Straightforward to minimize
based on experience with current generation of HxRG detector/amplifier
combinations.\\

\ref{subsec:sched} & Scheduling & \\ & HLS & Optimal to evenly space
observations over full time of survey, to extent permitted by other
requirements. Current example schedules vary in PM outcome by factor
of $\gtrsim 2$.\\

& EML survey & Programming an occasional larger dither will
significantly help calibrate for general astrometry. Largest possible
time-spacing between first and last exposures is optimal; regular
intermediate observations will increase understanding of long-term PSF
variations. \\

& GO & The TAC process should allow for multi-year GO proposals to
optimize PM baselines. For proposals covering large sky areas, time
between field revisits should be maximized. \\

\ref{subsec:jitter} & Jitter & This may be an issue for WFIRST where
it was not for \textit{HST}, given large requirement (14 mas). Requirements
of the HLS for galaxy shape determination should help. \\

\ref{subsec:dmg} & Data Management & Downloading every read with no
coadds for at least part of the FoV is highly desirable for spatial
scanning observations (see Section~\ref{sec:exoplanets}).  Downloads
of GS postage stamps are crucial and inexpensive for PSF
jitter correction. \\

\ref{subsec:archive} & High-level data products \& Archive &
Astrometry (linked to the \textit{Gaia} frame) and astrometric
uncertainties (including PSF centroiding error estimates) should be
part of the high-level products. A requirement should be set on the
astrometric uncertainty. The archive should allow for multiple
upgradable astrometric solutions.\\

\hline\hline
\end{tabular}
}
\caption{Summary of main recommendations for astrometry.}
\label{tbl:recs}
\end{table*}

\subsection{Geometric Distortion}
\label{subsec:geodist}

Geometric distortion (GD) is the most significant systematic
contributor for astrometry that is not currently covered by explicit
requirements for either the HLS or EML survey. A dedicated set of
observations to autocalibrate the GD of the WFI is currently being
considered. There are two main ways to solve for the GD:\ using
previous knowledge of the stellar positions in the field from existing
astrometric catalogs (the ``catalog'' method), or via autocalibration,
in which stellar positions themselves are iteratively solved for
together with the GD. Each of these approaches has different
advantages and disadvantages, but both depend strongly on the
precision with which the position of stars can be measured using
appropriate PSFs (e.g.,
Refs.~\citenum{2003PASP..115..113A,2004acs..rept....3A,2006A&A...454.1029A,
  2006acs..rept....1A,2009PASP..121.1419B,2010A&A...517A..34B,
  2011PASP..123..622B,2014A&A...563A..80L,2015MNRAS.450.1664L,
  2016A&A...595L...2M,2016ApJ...833..111D}).

The catalog method is less demanding of telescope time, since it
requires fewer images to calibrate the GD and monitor temporal
variations, but it strongly depends on the quality of the astrometric
catalog used as a reference, since the residuals and the systematic
errors present in the reference catalog \textbf{can propagate} in the GD
solution. In addition, unless the reference catalog is based itself on
images taken very close in time to the WFIRST calibration images,
PMs (and their errors) can introduce significant residuals
in the GD solution.
%
%
A recent technical report on the catalog method
(Ref.\ \citenum{bellini18}) also highlights the importance of using
accurate PSFs that take into account for jitter and inter-pixel
capacitance effects.

The autocalibration approach
requires more images, and therefore more telescope time, but offers a
self-consistent calibration solution and can be designed to be
formally insensitive to proper-motion-related errors.

On-sky GD calibration has historically been performed using large
dithered exposures (as wide as the FoV in some cases) of a
homogeneously-distributed, moderately dense stellar field. Stars in
the EML survey fields are homogeneously distributed, and while their
overall stellar density can be too high, this can be mitigated by
using only the brightest stars in each field to calibrate the GD. If
the exposure time has been carefully chosen, the bright, unsaturated
stars will still be reasonably far apart from each other, and their
surrounding neighbors typically a few magnitudes fainter, so that the
bright stars can still be considered fairly isolated.  The Baade
window was successfully used by Refs.~\citenum{2006A&A...454.1029A} and
\citenum{2014A&A...563A..80L} to calibrate the GD of two different ground-based, wide-field
IR detectors (ESO WFI\at MPG and HAWK-I\at VLT, respectively).  The
GC $\omega$~Cen, another possible target field for
WFIRST calibration, was used by Ref.~\citenum{2015MNRAS.450.1664L} to
calibrate the GD of the IR WFI VISTA InfraRed CAMera (VIRCAM) at the Visible and Infrared Survey Telescope for Astronomy (VISTA). The same
technique of using only the brightest stars could be applied to the
crowded regions in the core of $\omega$~Cen. On the other hand, the
stellar density near the tidal radius ($\sim 48^{\prime}$,
Ref.~\citenum{1996AJ....112.1487H}, 2010 edition) may be too low. Other
GCs could also be used for calibration, but because of
its overall high number of members and its large tidal radius,
$\omega$~Cen is the best target.

\begin{figure*}[t!]
\centering
\includegraphics[trim=80 0 0 0,clip=true,width=1.05\textwidth]{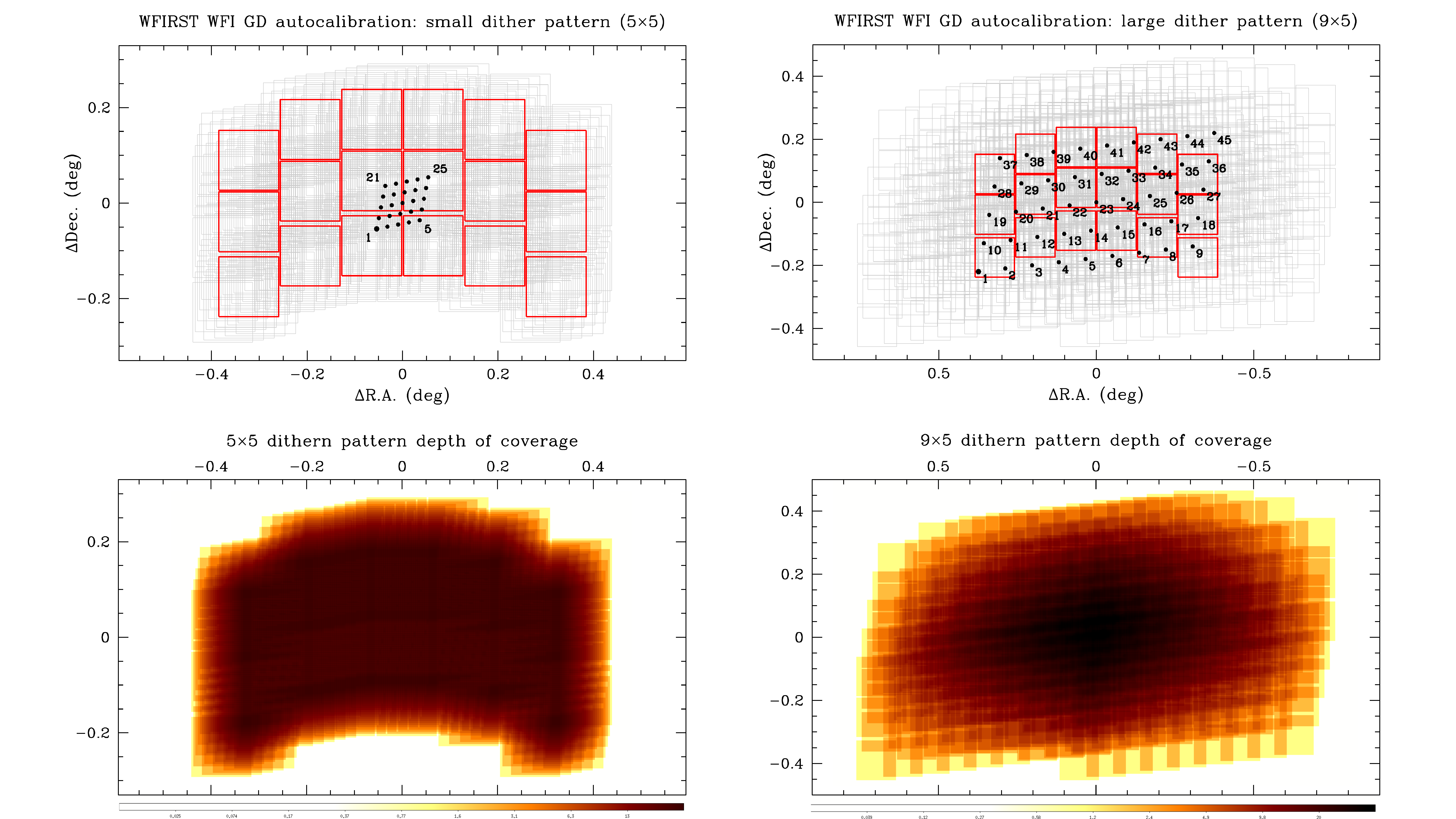}
\caption{\textbf{Left column:} Example of a small, $5\times 5$ dither
  pattern that covers each WFI detector from corner to corner.
  \textbf{Right column:} Example of a large, $9\times 5$ dither
  pattern that covers the entire FoV from corner to corner.
  \textbf{Top row:} Dither pattern layout on-sky, with the center of
  each dither marked by a black dot, the WFI outline of the central
  dither shown in red, and outlines of the other dithers in grey).
  \textbf{Bottom row:} Depth-of-coverage map (number of repeat
  observations as a function of position) for the assumed dither
  strategies, on a logarithmic scale. See \S \ref{subsec:geodist} for
  details.}
\label{f:gd}
\end{figure*}

\subsubsection{Example of an autocalibration strategy}

A possible autocalibration strategy for the WFIRST WFI could be
modeled on the calibration described in Ref.~\citenum{2015MNRAS.450.1664L} for the VIRCAM\at
VISTA detectors. The VIRCAM WFI comprises 16 2k$\times$2k VIRGO
detectors, for a total FoV of about 1.3$\times$1.0 sq. deg, but the
very large gaps between the chips bring down the effective FoV to 0.59
sq. deg. The calibration program (ESO proposal 488.L-0500(A),
  PI:\ Bellini) used a combination of small and large $5\times 5$
dithers. Large dithers were used to cover the gaps between chips,
monitor low-frequency distortions, and construct a single common
reference system for all observations; the small dithers were included
to enable independent modeling of the high-frequency residuals of the
GD within each chip (more in Ref.~\citenum{2015MNRAS.450.1664L}). The
choice of a $5\times 5$ dither pattern was a compromise to obtain a
sufficiently high-precision GD correction in a reasonable amount of
telescope time. Extremely small dithers (from a few subpixels to a few
pixels apart) are not strictly needed to characterize the PSF in
well-populated star fields, since nature distributes stars randomly
with respect to the pixel boundaries (see also
Ref.~\citenum{2000PASP..112.1360A}).

Figure \ref{f:gd} shows an example of a possible dither strategy for
WFIRST following this plan. In the left column is a plan for a small
$5\times 5$ dither pattern that covers each WFI detector from corner
to corner. In the top left panel, the black dots mark the center of
each of the 25 dithers, with the detector layout of the central dither
shown in red and other dithers in grey. The bottom left panel shows
the resulting depth-of-coverage map on a logarithmic scale, with a
maximum of 25 different images covering the same patch of sky. Most of
the map is covered by at least 22 images, but only 12--15 of these
come from the same chip, so that the same star is typically imaged in
12--15 different chip locations. The $5\times 5$ pattern never repeats
the same shift along the X or Y direction, thus guaranteeing that the
same stars will never fall on the same column or row, to minimize the
impact of possible detector defects or degeneracies in the distortion
solution.

The right column of Figure \ref{f:gd} shows an example plan for a $9\times 5$
pattern of large dithers that covers the entire WFI FoV from corner to
corner. Because of the rectangular shape of the WFI FoV, 9 dithers on
the X axis are needed to cover the FoV with similar spacing to the 5
dithers along the Y axis. As for the small dithers, the layouts and
centers of each pointing are shown on top and the resulting
depth-of-coverage map is shown on the bottom. In this case, the layout
results in a maximum of 39 different images at the center of the
pattern.

The dither patterns shown in Figure \ref{f:gd} all have the same
telescope rotation angle, but in order to properly calibrate the skew
terms of the distortion, a few observations of the same field at
different roll angles would be highly beneficial. It is not obvious to
suggest exactly how many of these rotated exposures should be taken,
but sampling the full circle every 45--60$^\circ$ should suffice.  The
total FoV covered by the large dither pattern in Figure \ref{f:gd}
allows for the central pointing to be rotated by any angle and still
be fully within the covered region.

The proposed dither strategy makes use of 25 small dithers and 45
large dithers for a given filter, plus 6 or 8 additional pointings (60$^\circ$ or
45$^\circ$ sampling, respectively) to constrain the skew terms, for a
total of 76--78 distinct pointings. Experience calibrating the
\textit{HST} GD shows that convergence in the GD solution is achieved
when stellar positions transformed from one image to another taken
with a different pointing have rms residuals comparable to the stellar
centroiding errors. Simulations to assess the precision of the GD
correction as a function of the adopted dither strategy are ongoing
(Bellini, in preparation) to determine, among other things, whether
fewer pointings than the example shown here could be sufficient.

An additional complication introduced by WFIRST's large FoV is due to
the use of tangent-plane projections:\ adopting the same projection
point for images taken more than a few arcminutes apart results in
significant positional transformation residuals. The GD calibration
therefore has to be carried out on the celestial sphere rather than on
any given tangent plane, adding an extra layer of complexity (see also
Ref.~\citenum{2015MNRAS.450.1664L}).

\subsubsection{Long-term monitoring of the GD solution}

The EML survey is intended to characterize the PSF and fine-tune the
GD solution of the WFIRST WFI, possibly including other sources of
systematic effects such as intra-pixel sensitivity variations. The
current design of the EML survey employs small dithers and fixed
rotation angle, suggesting that a satisfactory autocalibration of the
GD using only EML survey images will be very hard to achieve. The
catalog method could instead be used to fine-tune and monitor the GD
solution, probably using \textit{Gaia} as the reference catalog, but
the lack of different telescope rotation angles may result in poorly
constrained skew-term variations if these are present, as is the case
for the WFC of the Advanced Camera for Surveys on \textit{HST}. A
preliminary investigation into the possibility of using EML-like
simulated images of the Bulge, \textit{Gaia} absolute stellar
positions, and WebbPSF-based WFIRST PSF models (Bellini, in
preparation) showed that:\ (i) stellar positions measured by PSF
models that do not take into account jitter variations are
significantly affected by pixel-phase-like errors (of the order of a
few to several hundredths of a pixel); and (ii) the density of Gaia
stars in the simulated EML survey field (about 5000 stars per chip) is
adequate to solve for the GD.

Improved PSF models, either derived independently for each individual
exposure or as a function of jitter rms, will address the pixel-phase
issues and allow time-monitoring and fine-tuning of the GD solution.
Jitter will vary with the reaction wheel speed, particularly at speeds
that excite a structural vibration mode. Thus, the level of jitter is
expected to change with time even on short time scales, but should
have an rms well below 14 mas most of the time.  Regardless, because
of the time variability, and because excitation of the telescope
structure can cause line-of-sight jitter not sensed by the gyroscopes,
the guide-star data will be extremely valuable for characterizing the
jitter.  It would therefore be very useful to downlink the reads of
the guide windows together with each exposure, especially given the
small amount of additional data involved.  Furthermore, the
jitter-dependent PSF models obtained for the filters employed in the
EML survey will not necessarily apply equally well to other WFIRST
filters. If this turns out to be the case, it would be helpful to
include settling criteria that allow more stringent jitter rms
constraints when images are taken for the purpose of calibrating
and/or monitoring the GD. The current settling criteria include
constraints on both position and angular rates, but additional
criteria aimed at achieving better stability prior to calibrations
would be helpful and should be investigated, since a smaller jitter
rms implies smaller pixel-phase errors, helps in removing the
degeneracy between centroiding accuracy and GD residuals, and would
thereby make calibration more efficient.

Given that the EML survey will make use of only two of the WFI
filters, it is important to note that filter elements are known to add
significant contributions to the GD (e.g.,
Refs.~\citenum{2010A&A...517A..34B,2011PASP..123..622B,2014A&A...563A..80L}). Exposures
taken with the other filters must be used to monitor the time
dependency of the GD solution in those filters. In principle
\textit{all} WFI exposures can, and probably will, be used for this
purpose. This would typically be done with the catalog method, but
autocalibration can be applied when properly-dithered exposures are
available, so that any variation of the GD solution in all filters can
be mitigated to the fullest extent possible.

The possibility of ground calibration of both the distortion and
intra-pixel sensitivity variations (see \S \ref{subsec:subpix}) should
also be considered. As was found with \textit{HST}, a successful
astrometric calibration could reasonably be expected to improve
WFIRST's point-source localization, and therefore all
astrometry-related measurements, by an order of magnitude. Such an
improvement would multiply WFIRST's reach in distance or velocity
sensitivity for astrometry, thereby unlocking an entirely new space
for discoveries.

\subsection{Pixel-level effects}
\label{subsec:subpix}

\begin{figure*}
\includegraphics[width=\textwidth]{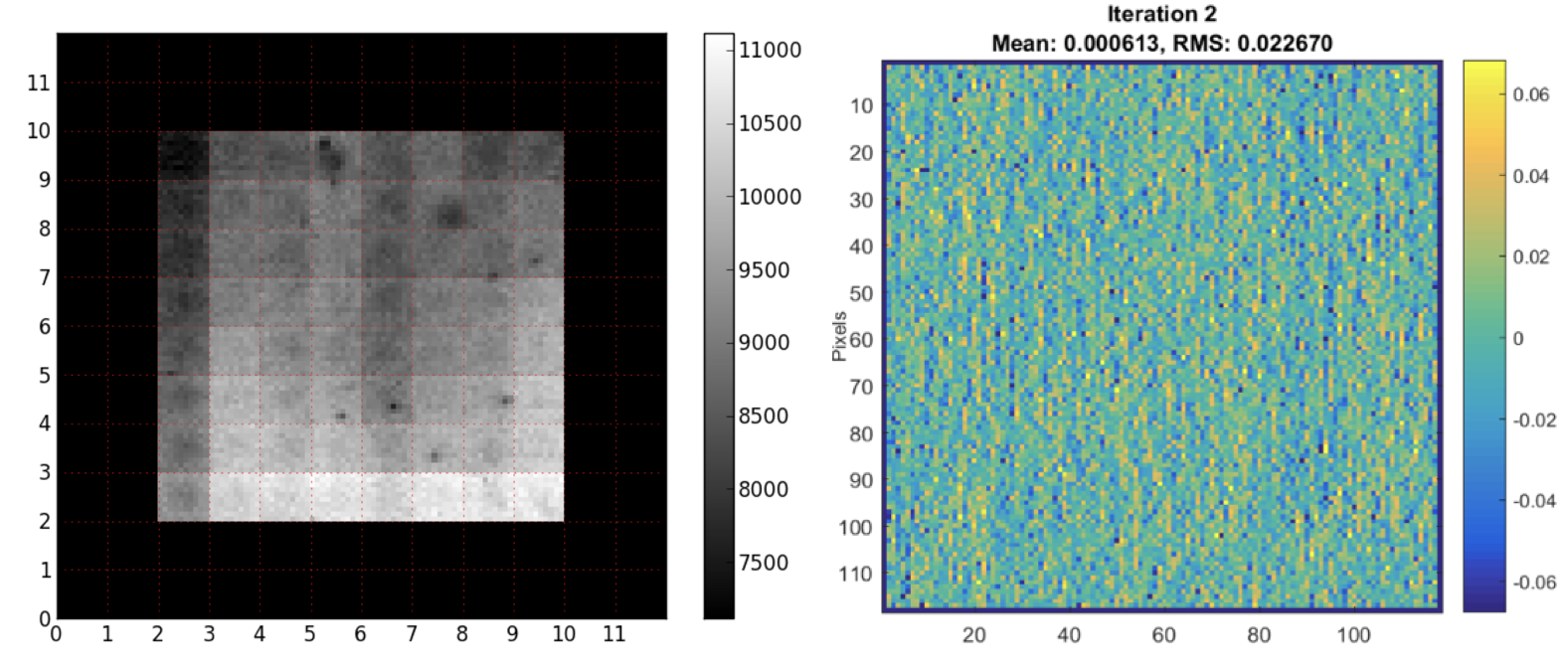} 
\caption{{\bf Left:} Figure 4 of Ref.~\citenum{hardy2014}, showing the pixel
  response at 650 nm for an 8-pixel-square region of an H2RG
  detector. {\bf Right:} Pixel offsets in a $128\times128$ region of
  an H2RG detector (Figure courtesy Michael Shao).}
\label{fig:pxpe}
\end{figure*}

\subsubsection{Quantum efficiency variations}

\label{subsec:QEvar}
Variations in the quantum efficiency (QE) \emph{within} a single pixel can
affect the accuracy of localization and therefore the astrometric
precision. Ref.~\citenum{hardy2014} measured the intra-pixel response
function for the H2RG detectors to be flown on JWST, which are direct
precursors of those planned for WFIRST. They found that the variations
in the response per pixel (shown in the left-hand panel of Figure
\ref{fig:pxpe}) appear to be mainly caused by redistribution of
charges from pixel to pixel rather than by variations in pixel
sensitivity. The most important effect in redistributing charge
between pixels was the diffusion of charge to neighboring pixels,
followed by interpixel capacitance (measured between 0 and 4 percent).
They also find that the type of small defect visible in Figure
\ref{fig:pxpe} occurs in roughly ten percent of pixels. Additional
testing of next-generation detectors more closely resembling those to
be used for WFIRST is ongoing, but we expect that they will exhibit
lower levels of variation.

\subsubsection{Placement error}
\label{subsec:pxpe}

To translate pixel-level effects into predictions for the precision of
localization, Michael Shao's group has made some preliminary
measurements of the pixel placement error in H2RG detectors.  The
``effective'' pixel position, which is defined as the location of the
centroid of the QE within each pixel, was
measured in these tests relative to an ideal coordinate system. Pixel offsets can have multiple causes,
including the QE variations within a pixel discussed in
Ref.~\citenum{hardy2014} and in \S\ref{subsec:QEvar}.  These tests considered
the pixel offset of a 128$\times$128 pixel section of a H2RG detector
and measured a rms $\sim$0.02 pixel offset error for a source measured in
a single exposure, twice the value
assumed in this work for single-exposure precision (see
\S\ref{ss:1.1}).  The right-hand panel of Figure \ref{fig:pxpe} shows
the pixel offset in the X direction for the portion of the H2RG
detector that was tested. By eye, the pixel placement errors appear to
be random, in which case relative astrometry for two stars falling within the 128$\times$128 region could be improved to better than 0.02 pixel by centroiding using the average of neighboring pixels. However, it is also possible that the neighboring 128$\times$128 group of pixels are systematically offset from the group tested (this type of systematic shift of a group of pixels has been seen in CCDs), in which case the relative astrometric error cannot be improved by averaging. The tested region is so far not large enough to detect this type of larger-scale systematic error. Currently, the accuracy of pixel
position measurement is estimated at $\lesssim$0.5\%, but the group is
in the process of more thoroughly verifying this as well as pushing
towards 0.1\% accuracy level. The main sources of error now
  are spurious fringes due to ghost reflections from, e.g., the dewar
  window, which have been minimized by tuning the laser over a broad
  range of wavelengths.

A 0.02-pixel position error corresponds to a single-image astrometric
error of about 2 mas, so calibration of this effect will be important
if WFIRST wants to deliver astrometry at the level of \textit{Gaia}
(10--100 times better) or even LSST precision (5--10 times
better). Multiple dithered images and spatial scanning can be used to
improve accuracy over the ``raw'' single-image error, depending on the
brightness of the targets.

\subsubsection{Ground- versus space-based calibration of subpixel effects}

A limitation of calibrating subpixel effects once the telescope is in
space is the issue of telescope jitter, which can change the PSF on a
time scale of hours. The use of images of crowded fields to solve for
subpixel errors in the detector relies on a stable PSF over a period
of time long enough to collect sufficient photons to calibrate
subpixel effects. The presence of time-variable telescope jitter
prevents this from happening by many orders of magnitude. It is almost
certain that the combination of jitter and photon noise will not allow
on orbit calibration better than just assuming a perfect detector,
given the measured 0.02 pixel errors in H2RG detectors.

It may not be as time-consuming to scan and calibrate this type of
variation on the ground as indicated in Ref.~\citenum{hardy2014}. They measured intrapixel QE variations by scanning a spot image
across $\sim$100 pixels using an extremely time-consuming process. For
larger regions containing $\sim10^4$ pixels, the accuracy of
measurements of pixel spacing using this approach will be limited by
the positional accuracy of the translation stage used to perform the
scan, which will likely be less accurate than the micrometer stage
used by Ref.~\citenum{hardy2014}. Therefore, while this is a good
approach to measure intrapixel QE, it's not sufficient for calibrating
the dimensional accuracy of a large focal plane array for
astrometry. The approach of scanning a spot across a pixel
individually would be prohibitively slow for the WFIRST focal plane,
which will contain 300 million pixels.  Instead, the tests described
in \S~\ref{subsec:pxpe}, which consider all pixels simultaneously,
could potentially be scaled up to calibrate all the detectors before
launch. Current estimates are that this scanning process can calibrate
the focal plane array roughly $10^4$--$10^5$ times faster than the
technique in Ref.~\citenum{hardy2014}. For a $\sim$300 megapixel camera
like the WFIRST WFI, such a calibration is estimated to take about
1--2 weeks, not including setup time.

\subsubsection{Persistence}
\label{subsec:persist}

Persistence of brightly illuminated regions is known to affect H2RG
devices, especially in areas that have been saturated beyond the
full-well depth (see the Wide Field Camera 3 Instrument
Handbook, \cite{wfc3-handbook}
section 7.9.4).  Characterizations of the persistence for both H2RGs
and H4RGs are currently ongoing in several laboratories, which should
establish a model for the persistence amplitude and decay time.  Such
a model can then be implemented to test the biases arising from
persistence, which are relevant not only for precision astrometry but
also for weak-lensing measurements in the HLS.  It remains to be
confirmed whether the model from ground-based testing is consistent
with the persistence experienced in flight, for which several
exposures of suitably bright stars should be sufficient.

If persistence is found to be problematic, a dark filter could be
employed to block the incoming light during slews, or slew
trajectories could be chosen to avoid bright stars.  Experience from
\textit{HST} indicates that without a dark filter, persistence during
slewing/tracking will be significant for stars brighter than 4th
magnitude at the maximum slew rate. Given that there is $\sim$1 star
brighter than 6th magnitude per 10 square degrees of sky, avoiding
these sparsely distributed sources during slewing should be fairly
straightforward.

\subsubsection{Brighter-fatter effect}
The brighter-fatter phenomenon is a well-known detector characteristic in CCDs whereby objects that are brighter have a larger PSF (i.e., are fatter; see e.g. Ref.~\citenum{2015JInst..10C5032G}).  This complicates many PSF-dependent measurements and characterizations, which generally operate under the assumption that the PSF size is invariant due to changes in flux or exposure time. More recently, this effect has been observed in an H2RG detector similar to the H4RG detectors planned for WFIRST\cite{2018SPIE10709E..1KR,2018PASP..130f5004P}.  There are currently laboratory efforts to understand and quantify this effect and corresponding efforts to software-based mitigation strategies. These efforts include experiment emulation for WFIRST at the Precision Projector Laboratory, a detector emulation and characterization facility designed to understand detector systematics at the demanding level required by weak lensing experiments\cite{2018arXiv180106599S}.  While we are at the early stages of quantitative studies of the brighter fatter effect on HxRG detectors, we expect there to be an induced astrometric bias, especially for undersampled point sources.  Depending on the flux and position of the source relative to the pixel grid, this effect could redistribute flux from the PSF center to neighboring pixels asymmetrically, causing the observed centroid to shift by up to 1\% of a pixel width, requiring mitigation procedures for sub-1\% astrometric measurements. 

\subsubsection{Mitigation strategies}

Spatial-scanning and diffraction-spike measurements
(\S~\ref{sec:exoplanets}) distribute the photons over hundreds or
thousands of pixels, and are therefore more robust against pixel-level
effects. Spatial scans are also robust against jitter
(\S~\ref{subsec:jitter}).  A complete summary of requirements for an
astrometric spatial scanning mode will be presented in an accompanying
report (Casertano et al., in preparation).

\begin{figure*}[t!]
\includegraphics[width=\textwidth]{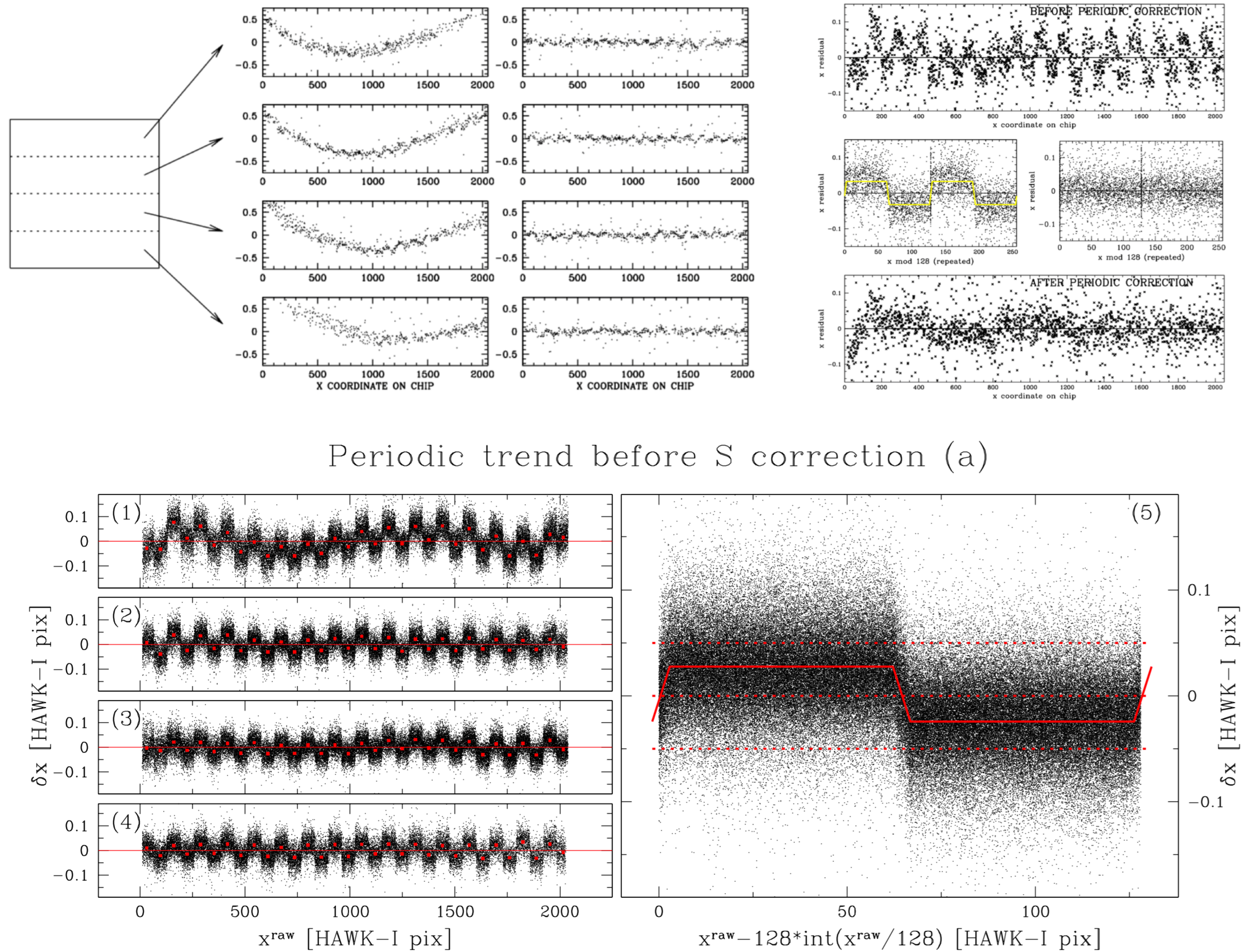}
\caption{Hysteresis in the HAWK-I detectors on the VLT, precursors to
  those planned for WFIRST.  \textbf{Top:} Results from an
  investigation by Anderson. Four horizontal strips of the detector
  were analysed, as shown on the far left. The central columns show
  the distortion as a function of $x$ position in each strip; the
  vertical scale is in pixels (1 pix = 100 mas). The raw distortion is
  shown to the left, while to the right are shown the residuals after
  subtraction of a smooth global polynomial, revealing a
  high-frequency periodic signal. On the far right are close-up views
  of the astrometric effect. The top row shows one
  polynomial-subtracted stripe; while the central row shows the
  residuals phase-wrapped by 128 columns. The resulting step function
  (left center row) has a half-amplitude of 0.035 pixel; the right
  center row shows residuals after subtraction. The unwrapped,
  corrected residuals are shown in the bottom row.  \textbf{Bottom:}
  Results from Ref.~\citenum{2014A&A...563A..80L}. The left-hand panel shows
  $\delta x$ residuals for each of the four HAWK-I detectors after the
  polynomial correction is applied. The right-hand panel shows a
  periodogram, with a period of 128 pixels, containing all the points
  plotted on the left, with the median shown as a solid red line. The
  dashed red lines are at 0 and $\pm 0.05$ HAWK-I pixels (about $\pm
  5.3$\,mas).}
\label{fig:hysteresis}
\end{figure*}

\subsection{Filters \& Color-dependent systematics}
\label{subsec:color}

Color-dependent systematic errors in the geometric
distortion  may
have a significant impact on WFIRST.  The bluest filters of the
WFC3/UVIS detector on
\textit{HST} have color-dependent residuals of a few hundredths of a
pixel (see Fig.~6 of Ref.~\citenum{2011PASP..123..622B}). In that case, it
is likely that the problem is due to a chromatic effect induced by
fused-silica CCD windows within the optical system, which refract blue
and red photons differently and have a sharp increase of the
refractive index below 4000\AA. As a consequence, the F225W, F275W,
and F336W filters are the most affected. WFIRST detectors are not
based on silica charge-coupled devices (CCDs), but the possibility of similar color-dependent 
systematic effects should be taken into account.

In addition, especially for wide-band filters, the geometric
distortion affecting redder and bluer photons is likely to be slightly
different due to diffraction and optical distortion. A test using WFC3/UVIS observations in
the F606W filter of blue-horizontal-branch and red-giant-branch stars
in $\omega$~Cen showed that the measured positions of blue and red
stars are off by $\sim$0.002 UVIS pixels on average, with respect to
their true positions. Filter-dependent residuals could therefore
introduce small but still significant color-dependent systematic
effects in wide-band WFIRST filters, including global (position-independent) effects. These filter-dependent
systematics are expected to be stable over time, so their calibration
should be straightforward.

\subsection{Hysteresis in readout electronics}
\label{subsec:hyster}

WFIRST should be aware that, on at least one existing instrument with
HxRG detectors, the readout electronics are affected by significant
hysteresis \cite{2014A&A...563A..80L}.  The effect was first
observed by J. Anderson in data taken by the High Acuity Wide field K-band Imager (HAWK-I) on the
Very Large Telescope (VLT) while observing a standard field to help calibrate JWST and the
HAWK-I detector itself, and later by Ref.~\citenum{2014A&A...563A..80L} in
all the fields observed with HAWK-I during the commissioning of the
detector (see their Table 2). Figure \ref{fig:hysteresis} shows the
results of Anderson's investigation on the top, and of
Ref.~\citenum{2014A&A...563A..80L} on the bottom. Plotted are the positional residuals of stars in the
       ``distortion-free'' master frame transformed into the raw
       reference system of the detector and the raw positions of the
       same stars in the raw reference system of the detector. The
       master frame was constructed iteratively using the
       auto-calibration method to solve for the geometric distortion. In both cases, a periodic
signal in the astrometric residuals is observed in the calibration
data, believed to be caused by the alternating readout directions for
the different amplifiers. A detailed description of the effects of
hysteresis on astrometry and how to minimize them is presented in
Section~5 of Ref.~\citenum{2014A&A...563A..80L}. For the H2RG chips of the
HAWK-I detectors, the hysteresis effect produces $\pm0.035$ pixels (or
$\pm$3.7 mas) of systematic error, that can be easily modeled and
corrected for. The WFIRST detectors will have 32 amplifiers, and
scientific full-frame images will make use of all of them.  If similar
hysteresis effects to those of the H2RG of the HAWK-I camera are
present as well in the WFIRST detectors, they will likely be easily
minimized as was done for HAWK-I.

\subsection{Scheduling}
\label{subsec:sched}

A small amount of extra attention to scheduling can give great payout
for astrometric measurements. As suggested by Equation
\ref{eq:PMAccuracy}, the ideal scheme for measuring PMs is to space revisits to the same field as evenly as possible over the longest possible time baseline, in the interest of minimizing both random and systematic errors. Maximizing the time baseline reduces the random error, while increasing the number of epochs protects against systematics, which are often time-dependent with unknown correlations. Here we discuss a few considerations
for the core science and guest observing components of the WFIRST
mission.

\subsubsection{High-Latitude Survey}
\label{subsec:sched:hls}

The observation of High-Latitude Survey (HLS) fields is planned to take place over a 5-year
baseline, but the detailed schedule is not yet finalized. Starting the
HLS with an initial exposure of each field in Year 1 and re-observing
at least once in each field as late as Year 5 would produce an
astrometric survey that extends \textit{Gaia}'s PM precision ($\sim$
25--50\,\uas\,yr$^{-1}$) to stars six magnitudes fainter than
\textit{Gaia} can reach.  However, the current range of schedules
considered for the HLS can affect this projected precision by factors
of a few. We recommend breaking ties between otherwise equivalent
programs by considering the time-distribution of revisits.

\subsubsection{Exoplanet MicroLensing Survey}
\label{subsec:sched:bmls}

The EML survey cycles through its 10 target fields at 52-second
intervals using the wide filter, with one set in the blue filter every
12 hours. Currently there are six total seasons planned, half in
spring and half in fall, in order to measure relative parallaxes with
a similar level of accuracy to that of PMs. Some observations will be
front-weighted at the start of the survey, but at least one season
should be planned for the end of the mission to obtain the longest
possible time baselines. Pointing and solar-panel orientation
requirements are likely to separate seasons by almost exactly 6
months, which may cause some complications in calibrating
time-dependent PSF effects (see \S\ref{subsec:geodist}).

\subsubsection{Guest Observing}

Much of the science described in \S~\ref{sec:science} will likely be
carried out through Guest Observer (GO) and Guest Investigator (GI) programs. It
is therefore crucial that the benefit of the astrometric requirements
for the two core science programs listed above be made available to
GO/GI programs as well (by providing the calibration information for
these programs to GOs/GIs and allowing for multiple astrometric
calibrations within the archive) in order to achieve the promised
precision. This need is discussed in depth in \S~\ref{subsec:archive}.

Since WFIRST is an IR telescope, it can provide astrometry for
regions of the Galaxy, such as the disk plane and bulge, that are
completely inaccessible to the current generation of optical
astrometric instruments. The EML survey will cover one such region,
but to take full advantage of WFIRST's astrometric capabilities, it
will be crucial to allow multi-year proposals from GOs in
order to optimize for PM baselines, as it is currently done for
\textit{HST}.

\subsection{Jitter}
\label{subsec:jitter}

The current WFIRST requirements impose a maximum jitter rms of $\leq
14$ mas, far larger than \textit{HST}'s jitter rms of
2--5\,mas. Jitter of the size allowed by WFIRST's requirements could
have a significant impact on the shape of the PSF. Tightening this limit would clearly result in better astrometry, but would also translate into a higher cost for some of the telescope components and is thus beyond the mandate of our working group. We have thus considered how to mitigate the effect of jitter at the level set by existing requirements.

Preliminary
simulations of WFIRST's geometric-distortion corrections based on
EML-like Bulge images\cite{bellini18}, which make use of
time-constant, \emph{spatially-variable} library PSF
models adapted from WebbPSF for WFIRST\cite{webbpsf-wfirst}, show
that there is significant degeneracy (at the 0.02--0.05 pixel level)
between the achievable geometric-distortion correction and pixel-phase
errors in stellar positions.

One way to break the degeneracy is to spatially perturb the library
PSF in each individual exposure, so as to tailor the library PSF to
the particular jitter status of each image. Because of the geometric
distortion, jitter-induced PSF variations are expected to affect the
PSF of different WFI detectors in a different way. Using this
workaround to calibrate the GD would require images
with roughly 20--40 thousand bright and isolated sources,
homogeneously distributed across the WFI FoV, in order to map local
PSF variations on scales of 500 pixels or so.

Another possibility would be to exploit the enormous number of images
in the EML survey to map PSF variations at different jitter rms
values, thereby creating a jitter-sampled set of spatially-variable
model PSFs. This technique will be limited by the use of only 2
filters for the EML survey, since it is unclear whether the models
generated by these two filters will apply equally well to the rest of
the filterset.

\subsection{Data management}
\label{subsec:dmg}

Both pointed and spatially scanned astrometric observations will be
constrained by the data downlink rate. For the WFIRST reference
mission at L2, the downlink rate is estimated to be about 1.3 TB
\unit{day}{-1}, barring addition of extra ground stations
\cite{2015arXiv150303757S}. Particularly for pointed astrometric
observations, this limits the number of reads per exposure that can be
downloaded. The current plan is to allow configuration of the options
for averaging and saving reads, similar to JWST. For astrometry, two
options are particularly important:
\begin{itemize}
\item Downloads of GS postage stamps are crucial and
  inexpensive (0.1\% overhead!) for PSF jitter correction.
\item The ability to download every read with no coaddition for at
  least part of the field is especially desirable for spatial scanning
  observations, as discussed in Section~\ref{sec:exoplanets}.
\end{itemize}

\subsection{High-level data products and archive}
\label{subsec:archive}

The combination of a wide FoV, the plentiful availability of
reference stars with extremely accurate astrometry from Gaia, and the
resolution and stability of a space-based platform makes it possible
for WFIRST to achieve extremely accurate absolute astrometry: better
than 100 \uas\ for essentially all WFC imaging products (see \S
\ref{sec:3} and Ref.~\citenum{bellini17b}).  Any field observed more
than once by the WFI is thus a potential astrometric field, providing
the community with a wealth of opportunities for high-precision
astrometry studies in many different domains, including many of the
science cases described in \S\ref{sec:science} of this
report. However, for this potential to be realized for all users
throughout the mission, \emph{we strongly recommend that the data
  processing pipeline and the archive incorporate from the outset the
  necessary elements to obtain, propagate, and maintain in practice
  the astrometric accuracy that the mission characteristics make
  possible in principle.}

Many of the features that make WFIRST an excellent astrometric
instrument, notably the requirements for excellent PSF modeling and
thermal stability, are already addressed by the core science programs
(the HLS and EML surveys). These requirements are summarized in Appendix \ref{appx:corescience}. 
Given their use of repeated visits to the same fields, both of
these programs also offer an opportunity to produce excellent
astrometry for all observed stars in their footprints with little
extra analysis. \emph{We recommend that derived PMs should
  be provided as part of the object catalogs for both surveys once
  multiple epochs have been observed.} In the HLS particularly, a
further cross-match of stars to the LSST catalog would extend the LSST
survey into the IR regime for the region covered by HLS, identify
variable stars that can be used as standard candles (particularly for
RR~Lyrae stars, as the period-luminosity relation is much tighter in
the IR), and allow for cross-validation of PMs.

Specifically, we recommend that the mission consider the possibility
of achieving the following goals:

\begin{enumerate}
\item The initial (a priori) astrometric information for each image
  should be based on GS positions known to Gaia accuracy,
  together with an accurate WFC GD model and the
  analysis of guide-star window data to extract accurate instantaneous
  GS positions.

\item The WFC GD model should be verified, and
  updated if needed, with sufficient frequency to maintain no worse
  than 100 \uas\ precision, on the basis of on-orbit data on the
  geometric stability of the WFC focal plane.

\item The pipeline to generate Level 2 products should include an a
  posteriori alignment step based on cross-matching sources found in
  each image with the Gaia catalog; this information should be
  incorporated as an alternate (preferred) astrometric solution in the
  image metadata.

\item The accurate astrometry thus determined should be propagated to
  Level 3 and Level 4 data products.

\item The Archive should have the ability to retain and distribute
  multiple astrometric solutions for each data product, together with
  their pedigree and uncertainty.  The Archive could also incorporate
  community-provided astrometric solutions if deemed useful.

\item The Data Management System should have the ability to update the
  astrometric information for higher-level products when the
  astrometric solution for the contributing data products is updated.
\end{enumerate}

These recommendations are based in part on our experience with
\textit{HST}.  The astrometric accuracy available for \textit{HST}
data early in the mission was originally limited by the quality of the
GS positions, therefore modest effort was placed into
improving other components of the astrometric fidelity, such as the
knowledge and time evolution of the relative positions of instruments
and guiders.  Now that substantially better positional accuracy is
available for guide and reference stars, retrofitting the \textit{HST}
pipeline and archive to improve the final absolute astrometric
accuracy of \textit{HST} processed data has proven complex and
resource-intensive.  We recognize that several of these
recommendations go beyond the current science requirements and may
exceed the baseline capabilities of the mission as currently planned.
However, incorporating these considerations, to the extent possible,
into the design of the WFIRST Data Processing and Archive systems will
greatly improve the quality and accessibility of mission data for
astrometry, improving science outcomes in this area and ultimately
reducing total development costs when compared with adding similar
capabilities at a later time.

~\\~\\
\noindent{\textbf{Acknowledgments.}}  This publication makes use of
data products from the Two Micron All Sky Survey, which is a joint
project of the University of Massachusetts and the Infrared Processing
and Analysis Center/California Institute of Technology, funded by the
National Aeronautics and Space Administration and the National Science
Foundation. This work has made use of data from the European Space
Agency (ESA) mission {\it Gaia}\cite{gaia-mainsite}, processed
by the {\it Gaia} Data Processing and Analysis Consortium (DPAC\cite{gaia-dpac}). Funding for the DPAC
has been provided by national institutions, in particular the
institutions participating in the {\it Gaia} Multilateral
Agreement.\\ RES was supported in part by an NSF Astronomy and Astrophysics
Postdoctoral Fellowship under grant AST-1400989.  The authors thank Charles Shapiro (JPL) for useful discussions, and the anonymous referees for comments that improved the paper.\\

\bibliography{references}

\renewcommand\cftfigafterpnum{\vskip12pt\par}
\renewcommand\cfttabafterpnum{\vskip12pt\par}

\listoffigures

\listoftables

\appendix

\section{Astrometry-relevant requirements for core science programs}
\label{appx:corescience}

Here we summarize the current state of relevant requirements for the 
two core-science surveys WFIRST will carry out, broken down into three categories: Basic Science
Requirements (BSRs), requirements for the High-Latitude Imaging Survey
(HLIS), and requirements for the Exoplanet MicroLensing (EML) survey.

\subsection{Astrometry with the High-Latitude Survey}
\label{sec:hls-reqs}

The current requirements related to astrometry, as of July 2017,
include the following:
\begin{description}
\item{{\bf BSR 2:} WFIRST WFI shall measure shapes of galaxies at
  z=0--2 in at least 2 bands and fluxes in at least 4 bands for
  photometric redshifts, at a depth equivalent to a 5-sigma point
  source detection at AB magnitude $J<26.9$ or $H<26.7$, with
  photometric accuracy of 1\% and with rms uncertainties (in the shape
  measurement filters only) below $10^{-3}$ in the PSF second moment
  and below $5 \times 10^{-4}$ in the PSF ellipticity, in the HLS
  imaging survey.}
\item{{\bf HLIS 7}: Obtain photometry, position, and shape
  measurements of galaxies in 3 filters ($J$, $H$, and F184), and
  photometry and position measurements in one additional color filter
  ($Y$; only for photo-z).}
\item{{\bf HLIS 8}: Obtain $S/N\geq 18$ (matched filter detection
  significance) per shape/color filter for galaxy effective radius $r_{\rm eff}=180$
  mas and AB mag = 24.7/24.6/24.1 ($J$/$H$/F184).}
\item{{\bf HLIS 9}: Determine PSF second moment to a relative error of
  $\leq 9.3\times 10^{-4}$ rms (shape/color filters only).}
\item{{\bf HLIS 10}: Determine PSF ellipticity to $\leq 4.7\times
  10^{-4}$ rms (shape/color filters only).}
\item{{\bf HLIS 11}: \textbf{The 50\% Encircled Energy (EE50) radius of
  the PSF} $\leq$ 0.12 ($Y$ band),
  0.12 ($J$), 0.14 ($H$), or 0.13 (F184) arcsec.}
%
\end{description}

The reference dither pattern for the HLS is a set of 3--4 dithers of a
size intended to cover the chip gaps, repeated to tile the field and
in each of 4 filters: Y, J, H, and F184. A second pass over each field
follows six months later at a different roll angle.

The projected, approximate bright and faint point-source limits of the
HLS are summarized in Table~\ref{tbl:hls-limits}. The faint limits are
as stated in the requirements above. The bright limits were estimated
based on the statement in the reference mission that pixels in the HLS
will be read non-destructively every 5.4 seconds, and assuming that
any pixels that saturate before the fourth such read will be
hard-saturated (see the discussion in
Ref.~\citenum{bellini17b}). Using the GS ETC, assuming 25\% of light
in the central pixel for all filters and 65,000 electrons as the
saturation level, the values listed in the table give the approximate
bright limit (probably accurate to within 0.3--0.4 magnitudes).

\begin{table}
\begin{center}
\begin{tabular}{lcccc}
\hline
\hline
HLS & $Y$ & $J$ & $H$ & $F184$ \\
 \hline
Bright limit & 15.5 & 15.6 & 15.7 & 15.3 \\
Faint limit & -- & 26.9 & 26.7 & -- \\
\hline
\hline
\end{tabular}
\end{center}
\caption{Bright and faint point-source limits for the WFIRST
  High-Latitude Survey (HLS; AB magnitudes), based on the requirements
  described in \S~\ref{sec:hls-reqs}.}
\label{tbl:hls-limits}
\end{table}

\subsection{Astrometry with the Exoplanet MicroLensing Survey}
\label{ss:4.2}

The current astrometry-related requirements being discussed (as of 29
June 2017) for the EML survey are:
\begin{description}
\item{{\bf EML 8:} Relative photometric measurements in the primary
  microlensing filter that have a statistical S/N of $\geq 100$ per
  exposure for a $H_{AB} = 21.6$ star.}
\item{{\bf EML 14:} The 50\% Encircled Energy (EE50) radius of
  the PSF in the wide filter shall be $<0$\farcs 15.}
\item{{\bf EML 19:} The relative astrometric measurements shall have a
  statistical precision of 1 mas per measurement for $H_{\rm
    AB}=21.6$.}
\item{{\bf EML 20:} Relative astrometric measurements will have
  systematic precision of 10 \uas\ over the full survey (stretch goal
  of 3 \uas).}
\end{description}

Currently, both narrow (2-pixel wide) and large (10$^{\prime\prime}$
wide) possibilities for dithering are being explored for this core
project. From the perspective of astrometric calibration, large
dithers are crucial to accurately measure skew and kurtosis in the
wings of the PSF.

Parallaxes and PMs over the bulge field are part of the mission of
this program to characterize the masses of the star-planet pairs that
will be discovered. The survey is therefore requesting the first two
(spring/fall) and last two bulge observing seasons over the full
time-baseline of the mission. This is also optimal for general
astrometry in the bulge fields but may pose problems for understanding
long-term variations in the PSF (on timescales of a year or so) prior
to the end of the mission.

\section{Typical astrometry-related queries to object catalog}

These queries were submitted to the Archive Working Group as part of
their ``20--questions'' use case development.

\begin{description}
\item[AWG-1] Give me positions, IR magnitudes, PMs, and associated uncertainties of all stars in the
  HLS or EML survey within a color-magnitude box/isochrone cutout.
\item[AWG-1a] Also return LSST optical magnitudes, PMs, and associated uncertainties for the
  selected stars.
\item[AWG-2] Return positions, magnitudes, PMs, distances, and associated uncertainties of stars in
  a specified field that LSST identifies as standard-candle variable
  stars (e.g. RR Lyrae).
\item[AWG-3] Run a group finder [or other analysis software] I provide
  on the above data.
\item[AWG-4] Give me all frames from any observing program, and any
  associated calibration information or data flags, that intersect a
  defined region on the sky. (I.e. it would be great to be able to
  re-reduce data from different observations to measure PMs with more
  frames/longer time baseline)
\item[AWG-5] Give me positions, parallaxes, PMs, and associated uncertainties of all \textit{Gaia} stars within a WFIRST pointing above a
  mag threshold.
\item[AWG-6] Extension of above: Give me predicted positions and
  uncertainties of \textit{Gaia} stars within a WFIRST pointing above
  a mag threshold at some observation date+time.
\item[AWG-7] Give me positions, PMs, magnitudes, and associated uncertainties for all XX-type
  stars within YY pc from the Sun in this ZZ WFIRST pointing.
\end{description}

\section{Glossary of Acronyms}
\textbf{2MASS}: Two-Micron All Sky Survey\\
\textbf{BSR}: Basic Science Requirements\\
\textbf{CCD}: charge-coupled device\\
\textbf{CMD}: color-magnitude diagram\\
\textbf{DM}: dark matter\\
\textbf{EE50}: 50\% Encircled Energy\\
\textbf{EML}: Exoplanet MicroLensing [survey]\\
\textbf{ESA}: European Space Agency\\
\textbf{ESO}: European Southern Observatory\\
\textbf{FoV}: field of view \\
\textbf{GC}: globular cluster\\
\textbf{GD}: geometric distortion\\
\textbf{GI}: Guest Investigator \\
\textbf{GO}: General Observer\\
\textbf{GS}: guide star\\
\textbf{HAWK-I}: High Acuity Wide field K-band Imager\\
\textbf{HBL}: hydrogen-burning limit\\
\textbf{HLS}: High-Latitude Survey\\
\textbf{HST}: Hubble Space Telescope\\
\textbf{HxRG}: ``H stands for HAWAII, an acronym for HgCdTe Astronomical Wide Area Infrared Imager. x denotes the number of 1024 (or 1K) pixel blocks in the x and y-dimensions of the array. R denotes reference pixels, and G denotes guide window capability.'' (see Ref.~\citenum{2011ASPC..437..383B})\\
\textbf{IR}: infrared\\
\textbf{JWST}: James Webb Space Telescope\\
\textbf{L2}: Lagrange point 2\\
\textbf{LG}: Local Group\\
\textbf{LSST}: Large Synoptic Survey Telescope\\
\textbf{MPG}: Max Planck Gesellschaft\\
\textbf{MPs}: multiple stellar populations\\
\textbf{MS}: main sequence\\
\textbf{MW}: Milky Way\\
\textbf{NASA}: National Aeronautics and Space Administration\\
\textbf{PM}: proper motion\\
\textbf{PSF}: point-spread function\\
\textbf{QE}: quantum efficiency\\
\textbf{UVIS}: Ultraviolet-VISible \\
\textbf{VIRCAM}: VISTA InfraRed CAMera\\
\textbf{VIRGO}: Raytheon HgCdTe 0.84-2.5 micron, 2048x2048 pixel IR detectors (see Ref. \citenum{2004SPIE.5499...68L})\\
\textbf{VISTA}: Visible and Infrared Survey Telescope for Astronomy\\
\textbf{VLT}: Very Large Telescope\\
\textbf{WFC}: Wide-Field Channel\\
\textbf{WFC3}: Wide-Field Camera 3\\
\textbf{WFI}: Wide-Field Imager\\
\textbf{WFIRST}: Wide-Field InfraRed Space Telescope \\

\end{document}